\documentstyle[aps,epsf]{revtex}

\def\be{\begin{equation}}
\def\ee{\end{equation}}
\def\bea{\begin{eqnarray}}
\def\eea{\end{eqnarray}}
\def\la{\langle}
\def\ra{\rangle}
\def\rd{\mbox{d}}
\def\ri{\mbox{i}}
\def\re{\mbox{e}}
\def\t12h{\frac{\theta_{12}}{2}}
\def\vac{|0\rangle}
\def\eps{\varepsilon}
\def\ep{\varepsilon^\prime}
\def\r#1{(\ref{#1})}
\def\nn{\nonumber\\}

\def\a{\alpha}

\def\PRB#1#2#3{{\sl Phys. Rev.} {\bf B#1} (#2) #3}
\def\PRL#1#2#3{{\sl Phys. Rev. Lett.} {\bf #1} (#2) #3}
\def\PLB#1#2#3{{\sl Phys. Lett.} {\bf #1B} (#2) #3}

\def\JETPL#1#2#3{{\sl Sov. Phys. JETP Lett.} {\bf #1} (#2) #3}
\def\JPCM#1#2#3{{\sl J. Phys. Cond. Matt.} {\bf #1} (#2) #3}
\def\IJMPA#1#2#3{{\sl Int. J. Mod. Phys.} {\bf A#1} (#2) #3}

\begin {document}

\title{\bf Quasi-1D spin-$\frac{1}{2}$ Heisenberg magnets in their
ordered phase: correlation functions}
\author {Fabian H.L. Essler$^{(a)}$, Alexei M. Tsvelik$^{(a)}$ and 
Gesualdo Delfino$^{(b)}$}
\address{$^{(a)}$ Department of Physics, Theoretical Physics,
        Oxford University\\ 1 Keble Road, Oxford OX1 3NP, United Kingdom}

\address{$^{(b)}$ Laboratoire de Physique Th\'eorique, Universit\'e de
Montpellier II\\ Pl. E. Batallion, 34095 Montpellier, France}

\maketitle
\begin{abstract}
\par
We study weakly coupled antiferromagnetic spin chains in their ordered
phase by combinining an exact solution of the single-chain problem with
an RPA analysis of the interchain interaction. 
A single chain is described by a quantum Sine-Gordon model and 
dynamical staggered susceptibilities are determined by employing the
formfactor approach to quantum correlation functions. 
We consider both antiferromagnetic order encountered in quasi-1D
materials like ${\rm KCuF_3}$ and spin-Peierls order as found in ${\rm
CuGeO_3}$. 
\end{abstract}


\par
\section{Introduction}

The increasing amount of neutron data on quasi-1D antiferromagnets
calls for creation of a theory capable of describing the behaviour of
such strongly anisotropic nonlinear systems in greater detail. Of
particular interest with regard to recent experiments on
${\rm CuGeO_3}$ and ${\rm KCuF_3}$ is the structure of the
multiparticle continua. In the present paper we discuss correlation
functions of three closely related problems involving
spin-$\frac{1}{2}$ Heisenberg chains: (i) a single chain with a
static alternating exchange, (ii) a single chain in a staggered
magnetic field, (iii) a system of weakly coupled spin-$\frac{1}{2}$
Heisenberg chains at low temperatures in the ordered phase. In the
latter case we consider two types of order: (a) spin-Peierls
\cite{has,pou,hir,reg} and (b) antiferromagnetic \cite{satija,welz,tenn}.  

An experimental realisation of situation (i) may be achieved in
(VO)$_2$P$_2$O$_7$. The analysis of the recent experiments conducted
by \cite{nagler} indicates that the strongest exchange occurs
through alternatively arranged molecules of two different types such
that the ratio of the exchange integrals is $J'/J \approx 0.722$.

The second situation (ii) may be realized when a Heisenberg magnet
with a staggered Land\'e factor is placed in an external magnetic
field. This is the case for copper benzoate \cite{dender} but the
difference in Land\'e factors is very small and for reasons that will
be discussed elsewhere the theory presented here does not apply. 

The spin-Peierls transition is a magnetoelastic transition which
occurs in quasi-1D antiferromagnets due to an effective four-spin
interaction. Such interactions may be generated by phonons
which modify the exchange integrals; however, the virtual absence of 
softening of the phonon spectrum in the spin-Peierls material 
CuGeO$_3$ \cite{hirota} suggests that there may be also other
mechanisms. The interaction between the staggered parts of energy
densities is strongly relevant and resolves in dimerization of the
lattice at a certain temperature T$_c$ and the formation of spectral
gaps in the spectrum of magnetic excitations. In any realistic system
phonons are three-dimensional which determines a three dimensional
nature of the spin-Peierls transition. A simple model with a phonon
mechanism has the following Hamiltonian
\bea
H &=& J\sum_{{\bf r},n}\left(1+(-1)^nu_{n,{\bf r}}\right)
\vec{\bf S}_{n, \bf{r}}\cdot\vec{\bf S}_{n+1,\bf{r}} 
+ \frac{1}{2}\sum_{n,\bf{r}\bf{r}'}u_{n,\bf{r}}
D({\bf{r} - \bf{r}'})u_{n,\bf{r}'}\nn
&+&\alpha J\sum_{{\bf r},n}\vec{\bf S}_{n, \bf{r}}\cdot\vec{\bf
S}_{n+2,\bf{r}} 
+\sum_{{\bf r},{\bf c},n}K({\bf{c}})\ \vec{\bf S}_{n, \bf{r}}
\cdot\vec{\bf S}_{n,\bf{r}+\bf{c}}  \label{ha}\ ,
\eea
where $n$ and $\bf{r}$ label lattice sites along and perpendicular to 
the chains and ${\bf c}$ are vectors connecting neighbouring
chains. In order to make contact with experiments we have included
n.n.n. intrachain as well as interchain magnetic interactions.

Since $J$ is usually much smaller than the phonon frequency, one can
neglect the kinetic energy of the $u$-field and consider $u$ as
commuting numbers. One can denote
\be
\epsilon(n, {\bf r}) = (-1)^n(\vec{\bf S}_{n,{\bf r}}\cdot\vec{\bf
S}_{n + 1,{\bf r}})
\ee
and integrate over the displacements. The result is 
\bea
H_{sP}&=& J\sum_{{\bf r},n}\vec{\bf S}_{n, \bf{r}}\cdot
\vec{\bf S}_{n+1,\bf{r}}+
\frac{1}{2}\sum_{n,\bf{r},\bf{r}'}\epsilon(n, {\bf r})J({\bf r}- {\bf
r}')\epsilon(n, {\bf r}')\nn
&+&\alpha J\sum_{{\bf r},n}\vec{\bf S}_{n,\bf{r}}\cdot\vec{\bf S}_{n+2,\bf{r}}
+\sum_{{\bf r},{\bf c},n}K({\bf{c}})\ \vec{\bf S}_{n, \bf{r}}
\cdot\vec{\bf S}_{n,\bf{r}+\bf{c}}\ ,
\label{hsp}
\eea
where $J({\bf r} - {\bf r}')$ is proportional to the matrix inverse of
$D$. Now one can forget about phonons and use (\ref{hsp}) as a general
model for the spin-Peierls state.

In their low temperature phases both $H_{sP}$ and the antiferromagnet
discussed below have order parameters. For the spin-Peierls model this
is 
the staggered energy density $\langle\epsilon(n,{\bf
r})\rangle=\epsilon_0$ (for simplicity we assume that there is no
antiferromagnetic order in the spin-Peierls model). Therefore in the
low temperature phase it is convenient to subtract from $\epsilon(n,
{\bf r})$ its average value ($:\epsilon: \equiv \epsilon -
\epsilon_0$). This leads to the following Hamiltonian
\bea
H_{sP} &=& \sum_{\bf r}H^{(0)}_{\bf r} +
\frac{1}{2}\sum_{n,\bf{r}\bf{r}'}:\epsilon(n, {\bf r}):J({\bf r}- {\bf
r}'):\epsilon(n, {\bf r}'): 
+\sum_{{\bf r},{\bf c},n}K({\bf{c}})\ \vec{\bf S}_{n, \bf{r}}
\cdot\vec{\bf S}_{n,\bf{r}+\bf{c}}\ ,\label{H1}\\
H^{(0)}_{\bf r} &=&  J\sum_{n}\vec{\bf S}_{n, \bf{r}}\cdot\vec{\bf S}_{n + 1,
\bf{r}} +\alpha J\sum_{n}\vec{\bf S}_{n, \bf{r}}\cdot\vec{\bf S}_{n+2,\bf{r}}
+ 2(J_a+J_b)\epsilon_0\sum_{n}\epsilon(n,{\bf r}) -
N\epsilon_0^2(J_a+J_b)\ ,
\label{sP}
\eea
where $J_{a,b}$ are the couplings (generated by the coupling to the
phonons) to the neighbouring chains in $a$ and $b$ directions (the
chain direction is chosen to be $c$). For definiteness we assume in
the following that $J_{a,b}<0$. However, the general case can be
treated by minor modifications of the formulas presented below.

Sometimes the purely one-dimensional model
(\ref{sP}) is used to describe the low temperature phase. We shall
show below that such approximation always becomes very poor close to
the spectral gap.

The second model we want to discuss consists of coupled spin-1/2
antiferromagnetic chains. Their Hamiltonian is
\be
H_{AFM} = J\sum_{n,{\bf r}}(\vec{\bf S}_{n,\bf{r}}
\cdot\vec{\bf S}_{n + 1,\bf{r}}) +J_\perp \sum_{n,{\bf r},{\bf a}} 
\vec{\bf S}_{n,\bf{r}}\cdot\vec{\bf S}_{n,\bf{r + a}}\ ,
\label{hb} 
\ee
where we take $J_\perp<0$ and {\bf a} are lattice vectors in
transverse directions. For simplicity we assume the transverse
coupling to be isotropic in $x$ and $y$ directions. Taking the
antiferromagnetic order (which we assume to be along the $z$
direction) into account by a mean-field analysis of the interchain
interaction one obtains the effective single-chain problem
\cite{schulz} 
\be
H_0 = J\sum_n\vec{\bf S}_n\cdot\vec{\bf S}_{n + 1} -
h\sum_n(-1)^nS_n^z-2NJ_\perp m_0^2\ , 
\label{stchain}
\ee
where $m_0=\langle(-1)^n S^z_n\rangle$ and $h=-4J_\perp m_0$. The
remaining interchain interactions will be treated in RPA (see Section
VI). 

\section{Transition to continuous description along the chains 
and bosonization}

We shall discuss the spin-Peierls case of weak dimerization 
$|\langle u\rangle|\ll 1$. In this case one can use a continuous
description of the spin-1/2 Heisenberg chain which, for $\alpha <
\alpha_c \approx 0.25$ \cite{zigzag}, is given by the
Gaussian model. In the framework of this model one can express spin
operators in terms of bosonic exponents (see {\sl e.g.}
\cite{book}). Thus, for instance, we have
\be
\epsilon(x)  = \frac{\lambda}{\pi a_0}\cos(\beta\Phi) + \mbox{less
singular terms}\ . \label{boson} 
\ee
The value $\beta = \sqrt{2\pi}$ was found by Nakano and Fukuyama
\cite{fukuyama} by using the Jordan-Wigner transformation with a
subsequent bosonization. There is a simplier way to establish this
value of $\beta$, namely, we can use the fact that the initial
Hamiltonian has an SU(2) symmetry which must be respected by the
bosonized form.
Replacing $u_n$ in Eq.(\ref{ha}) by the order parameter, that is
treating it as a number and using Eq.(\ref{boson}) we obtain the
sine-Gordon model. Since the SU(2) symmetry in the sine-Gordon model
is not present for general values of $\beta$ its requirement imposes a
restriction on the value of $\beta$. The corresponding point in the
sine-Gordon spectrum was discovered by Coleman \cite{coleman} and
Haldane \cite{haldane} who pointed out that at $\beta = \sqrt{2\pi}$
there are only two breathers; the first one has the same mass as
kink and antikink (let us call it $M$) and the second has the mass
equal to $\sqrt 3 M$. Therefore at $\beta = \sqrt{2\pi}$ kink,
antikink and the first breather realize an SU(2) triplet and the
second breather becomes an SU(2) singlet. 

In what follows we also need the bosonized form of the spin density in
the continuum ($\vec{S}_n\rightarrow a_0 \vec{S}(x)\ ,\ x=na_0$),
which is given by
\bea
\vec{\bf S} (x) &=& \vec{\bf J}_R (x) + \vec{\bf J}_L(x) + (-1)^n
\vec{\bf n} (x)\ ,\nn 
J^z_{R,L}&=&\frac{1}{2\sqrt{2\pi}}\left(\partial_x\Phi\mp\Pi\right)\
,\quad J^+_{R,L}=\frac{\mp i}{2\pi a_0} \exp\left(\mp
i\sqrt{2\pi}(\Phi\mp\Theta)\right),\nn 
n^z (x) &=& - \frac{\lambda}{\pi a_0} \sin \sqrt{2 \pi} \Phi (x), \quad
n^\pm (x) = \frac{\lambda}{\pi a_0} \exp [ \pm \ri \sqrt{2 \pi}
\Theta (x)].
\label{boso}
\eea
Here $n^a(x)$ are the components of the staggered magnetization,
$J_{R,L}^{\pm,z}$ are the currents of left and right moving fermions,
$na_0=x$ and $\lambda$ is a nonuniversal coefficient related to the
bandgap for the charge excitations in the itinerant electron model
that gives rise to the spin Hamiltonian. The field $\Theta$ is the dual of
the scalar field $\Phi$ and obeys $\partial_x\Theta(x)= \Pi(x)$, where
$\Pi$ is the canonical conjugate of $\Phi$. 
We note that (\ref{boso}) differ from the ``usual'' expressions
(see p. 270-271 of \cite{book}) by a shift of the bosonic field
by $\sqrt{\pi/8}$. This operation interchanges $\sin(\sqrt{2\pi}\Phi)$
and $\cos(\sqrt{2\pi}\Phi)$, but changes neither derivatives of $\Phi$
nor the dual field $\Theta$.

Notice that despite the fact that $\epsilon(x)$ has the same
dimension as the z-component of the staggered magnetization, 
it is given by a different operator (sine instead of cosine). In fact
$\epsilon(x)$ is the $2k_{\rm F}$-component of the charge density in the
system with a frozen charge field. 
As we shall see the situation is somewhat similar to that for the
spin-ladder (see for example \cite{ners}), but there are also certain
subtle differences. 
Substituting (\ref{boson}) into Eqs.(\ref{H1}) we get the
following  bosonized version of the spin-Peierls Hamiltonian
$H_{sP} = \sum_{\bf r}H_{\bf r}^{(0)}+ V_{phonon}+V_{spin}$
\bea
H_{\bf r}^{(0)} &=& \frac{v}{2}\int \rd x[\Pi_{\bf r}^2 +
(\partial_x\Phi_{\bf r})^2] + \mu\int\rd x\ \cos(\sqrt{2\pi}\Phi_{\bf r})
-N\epsilon_0^2 (J_a+J_b)\ ,\label{SG}\\ 
V_{phonon} &=&  \frac{\lambda^2}{2\pi^2a_0^2}\sum_{{\bf r},{\bf r}'}\int
\rd x:\cos(\sqrt{2\pi}\Phi_{\bf r}(x)):J({\bf r}- {\bf
r}'):\cos(\sqrt{2\pi}\Phi_{\bf r}'(x)):\label{vphonon}\\
V_{spin} &=&  \frac{\lambda^2}{\pi^2a_0^2}\sum_{{\bf r},{\bf c}}
\int\rd x\ K({\bf c})\ \vec{n}_{\bf r}(x)\cdot\vec{n}_{\bf r+c}(x)\ ,
\label{vspin}
\eea
where $\mu = \frac{2\epsilon_0\lambda}{\pi a_0}(J_a+J_b)$. Note that we
have kept only the most relevant terms and neglected {\sl e.g.} the
Umklapp term (for a discussion of the role of the Umklapp term see
\cite{fukuyama}). 
Our strategy is now to use exact results for
the sine-Gordon Hamiltonian (\ref{SG}) describing a single chain (see
also \cite{alexeiSP,schulzSP}) and to treat the residual interchain
interaction $V$ in the Random Phase Approximation (RPA).

The Hamiltonian (\ref{SG}) describing a single chain is of the form
$vH_0-N\epsilon_0^2 (J_a+J_b)$. The spectrum of $H_0$ is well-known
\cite{spectrum}: it is described in terms of a soliton, an
antisoliton, a bound state (called ``breather'') of mass $M$ and a
second breather of mass $\sqrt{3}{ M}$. The mass gap $M$ of the model
is due to the dimerization caused by the coupling to the phonons.
It is related to the scale $\mu$ as follows \cite{zam}
\be
\mu = \frac{\Gamma(\frac{1}{4})}{\pi\Gamma(\frac{3}{4})}\left[
M\sqrt{\frac{\pi}{4}}\frac{\Gamma(\frac{2}{3})}{\Gamma(\frac{1}{6})}
\right]^\frac{3}{2} .
\ee
The ground state energy density of $H_0$ is then given by
$e=-\frac{ M^2}{4}\tan\frac{\pi}{6}$, which in turn yields an
expression for the ground state energy density of $H$ as a function of
$\epsilon_0$. Minimization with respect to $\epsilon_0$ yields the
following mean field expressions for $\epsilon_0$ and mass gap ${\cal
M}$ of $H_{\bf r}^{(0)}$ as functions of the dimensionless (nonuniversal)
parameter $\frac{\lambda^2}{a_0}$ and the couplings $J_{a,b}$ and $J$ 
\bea
\epsilon_0&=&\sqrt{\frac{2}{\pi}}
\left(\frac{\tan\frac{\pi}{6}}{12}\right)^\frac{3}{2}
\left(\frac{2\Gamma(\frac{1}{6})}{\sqrt{\pi}\Gamma(\frac{2}{3})}\right)^3
\left(\frac{2\Gamma(\frac{3}{4})}{\Gamma(\frac{1}{4})}\right)^2
\frac{\lambda^2}{a_0}\left(\frac{|J_a+J_b|}{2J\kappa}\right)^\frac{1}{2}=: 
{\cal C}\left(\frac{|J_a+J_b|}{2J\kappa}\right)^\frac{1}{2}\ ,\nn
{\cal M}&=&v M=\frac{\tan\frac{\pi}{6}}{12}
\left(\frac{2\Gamma(\frac{1}{6})}{\sqrt{\pi}\Gamma(\frac{2}{3})}\right)^3
\left(\frac{2\Gamma(\frac{3}{4})}{\Gamma(\frac{1}{4})}\right)^2
\frac{\lambda^2}{a_0}\frac{|J_a+J_b|}{2}=:{\cal C}^\prime
\frac{|J_a+J_b|}{2}\ . 
\label{gap}
\eea
Here we have used $v=\frac{\pi}{2}Ja_0\kappa$ for the Fermi velocity,
where $\kappa$ is a function of the n.n.n. coupling $\alpha$. The
ratio of the constants ${\cal C}$ and ${\cal C}^\prime$ is found to be
\be
\frac{{\cal C}}{{\cal C}^\prime} = \frac{1}{3^\frac{3}{4}\sqrt{2\pi}}
\approx 0.175013\ .
\label{ratio}
\ee
Equation \ref{gap} makes it clear that the gap originates from the
interchain interactions.

\section{Sine-Gordon correlation functions at $\beta^2 = 2\pi$}
 
In this section we derive exact results for various correlation
functions of the sine-Gordon model (\ref{SG}) for $\beta =\sqrt{2\pi}$.
We start by constructing a convenient basis of states for the
sine-Gordon theory by means of the Zamolodchikov-Faddeev algebra. This
is based on the knowledge of the exact spectrum and scattering matrix
of the model \cite{zamo,karo}. We then formulate the problem of
calculating correlation functions in terms of {\sl formfactors} and
finally give explicit results for the first few terms in the
formfactor expansion.

The Zamolodchikov-Faddeev (ZF) algebra for the sine-Gordon model with
$\beta^2 = 2\pi$ was derived by Affleck \cite{affleck} who suggested
a representation which manifestly respects the SU(2) symmetry. 
As mentioned above there are three single particle states with mass
$M$ which form a triplet under the $SU(2)$ symmetry. The corresponding
creation and annihilation operators are denoted by $Z^+_a(\theta),
Z_a(\theta)$ ($a = \pm\frac{1}{2},1$). Here $1$ denotes the breather
state and $\pm\frac{1}{2}$ denote soliton and antisoliton states
respectively. In addition there is one single-particle breather state
with mass $\sqrt{3}M$, which transforms as a singlet under
$SU(2)$. Its creation and annihilation operators are denoted by
$Z_2^+(\theta), Z_2(\theta)$. 
As usual the eigenstates are parametrized by a rapidity variable
$\theta$ such that their momentum and energy are equal to
\be
p_j = M_j\sinh\theta_j, \: \epsilon_j = M_j\cosh\theta_j\ ,\qquad
\ee
where $M_j = \sqrt 3 M$ for the singlet state and $M_j = M$ for the
triplet states. By definition the ZF operators (and their hermitean
conjugates) satisfy the following algebra
\bea
{Z}_a(\theta_1){Z}_b(\theta_2) &=& S_{a,b}(\theta_1 -
\theta_2){Z}_b(\theta_2){Z}_a(\theta_1)\ ,\quad a,b=\pm\frac{1}{2},1\nn
{Z}_a(\theta_1)Z_2(\theta_2) &=& S_{a,2}(\theta_1 -
\theta_2)Z_2(\theta_2){ Z}_a(\theta_1)\ ,\quad a=\pm\frac{1}{2},1\nn
Z_2(\theta_1)Z_2(\theta_2) &=&
S_{2,2}(\theta_1-\theta_2)B(\theta_2)B(\theta_1)\ , 
\label{fz1}
\eea
where the two-particle scattering matrices $S_{ij}(\theta)$ are given
by
\bea
S_{a,b}(\theta)&=&\frac{\sinh\theta+\ri\sin\frac{\pi}{3}}
{\sinh\theta-\ri\sin\frac{\pi}{3}}=: S_0(\theta)\ ,\quad a,b=\pm\frac{1}{2},1\nn
S_{a,2}(\theta)&=&S_0(\theta+\ri\frac{\pi}{6})\
S_0(\theta-\ri\frac{\pi}{6})\ ,\quad a=\pm\frac{1}{2},1\nn
S_{2,2}(\theta)&=& \left(\frac{\sinh\theta
+\ri\sin\frac{\pi}{3}}{\sinh\theta-\ri\sin\frac{\pi}{3}}\right)^3\ .
\label{smat}
\eea
For the creation and annihilation operators we have 
\bea
{ Z}_a(\theta_1){ Z}_b^+(\theta_2) &=& S_{0}(\theta_2 -
\theta_1){ Z}_b^+(\theta_2){ Z}_a(\theta_1) 
+ 2\pi\delta_{ab}\delta(\theta_1 - \theta_2)
\ ,\quad a,b=\pm\frac{1}{2},1\nn
{ Z}_a(\theta_1)Z_2^+(\theta_2) &=& S_{a,2}(\theta_2 -
\theta_1)Z_2^+(\theta_2){ Z}_a(\theta_1)\ ,\nonumber\\
Z_2(\theta_1)Z_2^+(\theta_2) &=& S_{2,2}(\theta_2 -
\theta_1)Z_2^+(\theta_2)Z_2(\theta_1) + 2\pi\delta(\theta_1 - \theta_2)\ .
\label{fz2}
\eea

 From (\ref{smat}) it follows that $S_{i,i}(0) = - 1$ and $S_{i,i}(\infty)
= + 1$. Therefore particles with close momenta behave like free
fermions and particles far apart in momentum space behave like free bosons. 

We note that the soliton S-matrix $S_0(\theta)$ has simple poles at
$\theta=i\frac{\pi}{3}$ and $\theta= i\frac{2\pi}{3}$. These poles
correspond to the two breather bound states. In other words a soliton
and an antisoliton can form a bound state of either mass $M$ or mass
$\sqrt{3}M$. Although the S-matrix of light breathers is the same only
the pole $\theta=i\frac{\pi}{3}$ corresponds to a bound state -- the
heavy breather. The pole at $\theta= i\frac{2\pi}{3}$ is {\sl
redundant} \cite{zamo2}. The soliton-breather S-matrices
$S_{a,1}(\theta)$ and $S_{a,2}(\theta)$ exhibit soliton poles at
$\theta=i\frac{\pi}{2}\pm i\frac{\pi}{6}$ and
$\theta=i\frac{\pi}{2}\pm i\frac{\pi}{3}$ respectively.
All other poles are redundant and do not indicate the presence of
bound states.
>From the analytic properties of the S-matrices (\ref{smat}) we
deduce relations between the ZF operators. For example we find
\be
Z_2(\frac{\theta_1+\theta_2}{2})=\lim_{\theta_1-\theta_2\to
\frac{i\pi}{3}} { Z}_1(\theta_2){ Z}_1(\theta_1)\ .
\ee
Such relations between the ZF operators play an important role in what
follows.

States in the Fock space are constructed by acting with the operators
$Z_\eps^\dagger(\theta)$ on the vacuum state $\vac$
\be
|\theta_n\ldots\theta_1\rangle_{\eps_n\ldots\eps_1} = 
Z^\dagger_{\eps_n}(\theta_n)\ldots Z^\dagger_{\eps_1}(\theta_1)\vac ,
\label{states}
\ee
where $\eps_j=\pm\frac{1}{2},1,2$. We note that \r{fz1} together with
\r{states} implies that states with different ordering of two
rapidities {\sl and} indices $\eps_i$ are related by multiplication
with 2-particle S-matrices
\be
|\theta_n\ldots\theta_k\theta_{k+1}\ldots\theta_1
\rangle_{\eps_n\ldots\eps_k\eps_{k+1}\ldots\eps_1} = 
S_{\eps_k,\eps_{k+1}}(\theta_k-\theta_{k+1})
|\theta_n\ldots\theta_{k+1}\theta_k\ldots\theta_1
\rangle_{\eps_n\ldots\eps_{k+1}\eps_k\ldots\eps_1}
\label{order}
\ee

The resolution of the identity is given by
\be
1\!\!1=\sum_{n=0}^\infty\sum_{\eps_i}\int
\frac{\rd\theta_1\ldots\rd\theta_n}{(2\pi)^nn!}
|\theta_n\ldots\theta_1\rangle_{\eps_n\ldots\eps_1}
{}^{\eps_1\ldots\eps_n}\langle\theta_1\ldots\theta_n|\ .
\label{id}
\ee
The formfactor approach is based on the idea of inserting \r{id}
between the operators in a correlation function
\be
\langle {\cal O}(x,t){\cal O}^\dagger(0,0)\rangle
=\sum_{n=0}^\infty\sum_{\eps_i}\int
\frac{\rd\theta_1\ldots\rd\theta_n}{(2\pi)^nn!}
\exp\left({i\sum_{j=1}^n p_jx-\epsilon_jt}\right)
|\langle 0| {\cal
O}(0,0)|\theta_n\ldots\theta_1\rangle_{\eps_n\ldots\eps_1}|^2, 
\ee
and then determining the formfactors 
\be
F^{\cal O}(\theta_1\ldots\theta_n)_{\eps_1\ldots\eps_n}:=
\langle 0| {\cal
O}(0,0)|\theta_n\ldots\theta_1\rangle_{\eps_n\ldots\eps_1} 
\ee
by taking advantage of their known analytic properties.

 From a physical point of view we are interested in the Fourier
transforms  of the connected retarded 2-point correlators of
$\cos\sqrt{2\pi}\Phi$ and $\sin\sqrt{2\pi}\Phi$. Their formfactor
expansions are of the form 
\bea
D^{cos}(\omega,q)&=& \int_{-\infty}^\infty\rd x\int_0^\infty\rd t
\ e^{i (\omega+i\epsilon)t-i\frac{vq}{a_0}x} \langle
[\cos\sqrt{2\pi}\Phi(t,x) , \cos\sqrt{2\pi}\Phi(0,0)]\rangle \nn
&=& -2\pi\sum_{n=0}^\infty\sum_{\eps_i}\int
\frac{\rd\theta_1\ldots\rd\theta_n}{(2\pi)^nn!}
|F^{cos}(\theta_1\ldots\theta_n)_{\eps_1\ldots\eps_n}|^2
\left\lbrace\frac{\delta(\frac{vq}{a_0}-\sum_jM_j\sinh\theta_j)}{\omega - \sum_j
M_j\cosh\theta_j +i\epsilon}-
\frac{\delta(\frac{vq}{a_0}+\sum_jM_j\sinh\theta_j)}{\omega + \sum_j
M_j\cosh\theta_j +i\epsilon}\right\rbrace .\nn
\label{dcos}
\eea
Here we have reinserted the Fermi velocity $v$ and lattice spacing
$a_0$. The Fourier transform $D^{sin}(\omega,q)$ of the connected
retarded 2-point correlator of $\sin\sqrt{2\pi}\Phi$ is the dynamical
staggered susceptibility and will also be denoted by
$\chi^{\prime\prime}(\omega,q)$.

In order to implement the formfactor expansion it is very useful to
note that (like for general values of $\beta$) operators from different
representations behave differently under the charge conjugation
transformation
\bea
C\Phi C^{-1} &=& - \Phi\ ,\nonumber\\ C {
Z}_{\pm\frac{1}{2}}(\theta)C^{-1} &=& {\cal
Z}_{\pm\frac{1}{2}}(\theta), \qquad C { Z}_1(\theta)C^{-1} = - {\cal
Z}_{1}(\theta)\ ,\nonumber\\ C Z_2(\theta)C^{-1} &=& Z_2(\theta)\ . 
\label{chargeconj}
\eea
These transformation properties imply the following expansion
\bea
&&\sin[\sqrt{2\pi}\Phi(\tau, x)]|0\rangle = 
F_1\int\frac{\rd\theta}{2\pi}\re^{ - iM(t\cosh\theta -x\sinh\theta)}
{Z}_1^\dagger(\theta)|0\rangle \nn
&&\qquad+\int\frac{\rd\theta_1}{2\pi}\frac{\rd\theta_2}{2\pi}
\re^{ -iM[t(\cosh\theta_1 + \cosh\theta_2)-x(\sinh\theta_1 + \sinh\theta_2)]}
\ \times\nn
&&\qquad\qquad\qquad\times\
U(\theta_1,\theta_2)[{ Z}_\frac{1}{2}^+(\theta_1)
{ Z}_{-\frac{1}{2}}^+(\theta_2) - { Z}_{-\frac{1}{2}}^+(\theta_1){ Z}_{
\frac{1}{2}}^+(\theta_2)] + ...
\eea 
where $U(\theta_1,\theta_2) = U(\theta_2,\theta_1)$. Significantly,
due to the SU(2) symmetry transverse components of the staggered
magnetization have the same correlation functions as $n^z_{stag}$
(this is clear from the SU(2) symmetry of the Hamiltonian
(\ref{sP})). 
Consequently we conclude that Sine-Gordon 2-point
correlation functions (for $\beta=\sqrt{2\pi}$) of
$\sin\sqrt{2\pi}\theta$ and $\cos\sqrt{2\pi}\theta$ are the same as
2-point correlators of $\sin\sqrt{2\pi}\Phi$. 

The current operator $\partial_x\Phi$ is also odd in $\Phi$ and therefore
its expansion must begin with ${\cal Z}_0$ such that at small $q$ we
have
\be
\la\la \vec{\bf S}(\omega, q)\vec{\bf S}(-\omega, - q)\rangle\rangle \sim
\frac{\frac{v^2}{a_0^2}q^2}{\omega^2 - \frac{v^2}{a_0^2}q^2 - M^2} + ... 
\label{small}
\ee
where dots denote terms which have nonzero imaginary parts at higher
energies.  

As we will see the threshold of the dynamical spin susceptibility is
equal to $M$ for both $q = 0$ and $q = \pi$.
This is a distinct feature of the alternating chain. It is related to
the fact that kink and antikink create a bound state of the same mass. 
Recall that for the ladder chain (or S = 1 antiferromagnet for this
matter) where particles do not have bound states, the value of the
energy threshold at $q = 0$ is twice that at $q = \pi$ (see
Ref.\cite{ners}).

At frequencies smaller than $(1 + \sqrt 3)M$ the only
contributions to the imaginary part of the magnetic susceptibility
come from the first breather and kink-antikink pairs. 
Kink-antikink form factors can be calculated in the Sine-Gordon model
(for any value of the coupling $\beta$) along the following lines
\cite{karowski,smirnov}. Let us denote by $S_+(\theta)$
($S_-(\theta)$) the $S$-matrix eigenvalue corresponding to positive
(negative) $C$-parity obtained by diagonalising the kink-antikink
scattering: 
\bea
S_+=\frac{\sinh\frac{\pi}{2\xi}(\theta+\ri\pi)}
{\sinh\frac{\pi}{2\xi}(\theta-\ri\pi)}S_0(\theta)\ ,\qquad
S_-=\frac{\cosh\frac{\pi}{2\xi}(\theta+\ri\pi)}
{\cosh\frac{\pi}{2\xi}(\theta- \ri\pi)} S_0(\theta)\ , 
\eea
where $\xi = \frac{\pi\beta^2}{8\pi-\beta^2}$ and
\be
S_0(\theta) =- \exp(-i\int_0^\infty \frac{\rd x}{x}\frac{\sin\theta x
\sinh\frac{\pi-\xi}{2}x}{\cosh\frac{\pi x}{2} \sinh\frac{\xi x}{2}})\ .
\ee
Then, general
unitarity and crossing arguments imply that the corresponding kink-antikink 
form factors $F_\pm(\theta)$ are solutions of the following system of 
functional equations
\bea
F_\pm(-\theta)&=&S_\pm(\theta)F_\pm(\theta)\label{23}\\
F_\pm(\theta-2\ri\pi)&=&\pm F_\pm(-\theta)\,\,\,.\label{24}
\label{2partfun}
\eea
The ``minimal'' solutions of these equations are 
\bea
F_+(\theta)&=&\frac{\sinh\theta}{\sinh(\theta+\ri\pi)\frac{\pi}{2\xi}}
F_0(\theta)\nonumber\\
F_-(\theta)&=&\frac{\sinh\theta}{\cosh(\theta+\ri\pi)\frac{\pi}{2\xi}}
F_0(\theta)\ ,
\eea
where $F_0$ is given by
\be
F_0(\theta) =\sinh\frac{\theta}{2}\exp(\int_0^\infty\frac{dx}{x} 
\frac{\sinh\frac{x}{2}(1-\frac{\xi}{\pi}) \sin^2\frac{x(i\pi+\theta)}{2\pi}}
{\sinh\frac{x\xi}{2\pi}\cosh\frac{x}{2}\sinh x})\,\,\,.
\nonumber
\ee
By minimal solution we mean a solution 
containing only the expected bound state poles in the physical strip and with 
the mildest asymptotic behaviour at infinity. This prescription determines the 
minimal solution {\it uniquely}. An infinite number of non-minimal
solutions corresponding to all operators in the theory which are local
with respect to the solitons are
obtained multiplying the minimal solution by an analytic function of
$\cosh\theta$. However, if we require the form factor to be power
bounded in the momenta and to have only the bound state poles, we  
conclude that we can actually multiply the minimal solution only by a 
polynomial in $\cosh\theta$. For a given operator, it is possible to
put strong constraints on the asymptotic behaviour of its form
factors, and then on the degree of the allowed polynomial \cite{delmus}. 
In the sine-Gordon model this procedure is complicated by a
non-trivial behaviour of correlators in the ultraviolet
limit. Nevertheless, the result is that for the operators cos and sin
the allowed polynomial is of the zero degree, which means that their
form factors coincide with the minimal ones. The same conlusion can be
reached in a simpler way going to the free fermion point $\xi=\pi$,
where the form factors of sin and cos can be easily computed
remembering that
\bea
\cos\beta\Phi \sim  {\bar \Psi}\Psi, \:    
\varepsilon_{\mu\nu}\partial^\nu\Phi \sim J_\mu\,\,,
\eea
and that the sin is related to the elementary field by the equation of motion.

For the operators $\cos\beta\Phi$ and $\sin\beta\Phi$, at the specific value
of the coupling we are interested in, we find
($\theta_{12}=\theta_1-\theta_2$) 
\bea
\la 0|\sin\sqrt{2\pi}\Phi|\theta_1,\theta_2\ra_{-+} = \sqrt{3}(2d)Z^{1/2}
\frac{\cosh \theta_{12}/2}{\sinh 3\theta_{12}/2}\zeta(\theta_{12})
\label{sin} =F^{sin}(\theta)_{-+}\\
\la 0|\cos\sqrt{2\pi}\Phi|\theta_1,\theta_2\ra_{-+} = \ri\sqrt{3}
(2d)Z^{1/2} \frac{\cosh\theta_{12}/2}{\cosh
3\theta_{12}/2}\zeta(\theta_{12}) \label{cos}=F^{cos}(\theta_{12})_{-+}\\ 
\zeta(\theta) =c\sinh\theta/2 \exp\left\{2\int_0^{\infty}\rd
x\frac{\sin^2[x(\theta + \ri\pi)/2]\cosh\pi x/6}{x\sinh\pi x\cosh\pi
x/2}\right\} \nn
c=(12)^\frac{1}{4} \exp\left\{\frac{1}{2}\int_0^{\infty}\rd x
\frac{\sinh x/2\cosh x/6}{x\cosh^2 x/2}\right\}\approx 3.494607\
,\quad d=\frac{3}{2\pi c}\approx 0.136629\ ,
\eea
where the relative normalisation between the two operators can be fixed 
exploiting the asymptotic factorisation of form factors discussed in Ref.\,
\cite{cluster}. We note that $\zeta(\theta)$ is to be analytically
continued using the relation 
\be
\zeta(\theta) S_0(\theta) = \zeta(-\theta)\ .
\ee 
The additional factors $d$ and $c$ in \r{sin} and \r{cos} have been
introduced in order to simplify the reduction of multiparticle
formfactors using the annihilation-pole condition (for soliton
formfactors)
\bea
&&i{\rm Res}F^{\cal
O}(\theta_1\ldots\theta_{2n})_{\eps_1\ldots\eps_{2n}} 
\bigg|_{\theta_{2n}-\theta_{2n-1}=i\pi} = 
F^{\cal O}(\theta_1\ldots\theta_{2n-2})_{\ep_1\ldots\ep_{2n-2}}
\delta_{\eps_{2n}}^{-\ep_{2n-1}}\nn
\times && \left\{
\delta_{\eps_1}^{\ep_1}\ldots\delta_{\eps_{2n-1}}^{\ep_{2n-1}} - 
S^{\ep_{2n-1},\ep_1}_{\tau_1,\eps_1}(\theta_{2n-1}-\theta_1)\ldots
S^{\tau_{2n-3},\ep_{2n-2}}_{\eps_{2n-1},\eps_{2n-2}}
(\theta_{2n-1}-\theta_{2n-2})\right\}\ ,
\label{annpole}
\eea
where $S^{\eps,\eps}_{\eps,\eps}(\theta) =
S^{\eps,\ep}_{\eps,\ep}(\theta)=S_{1,1}(\theta)$ and all other
components are zero.
Multiparticle formfactors are discussed in some detail in Appendix
{\bf A}.

The form factor (\ref{sin}) has a pole at $\theta_{12} = -2\ri\pi/3$ 
corresponding to formation of a bound state - the first breather. The
breather formfactor $F_1$ is given by the residue of (\ref{sin})
divided by the three-particle coupling:
\be
|F_1|^2 = \frac{3^\frac{3}{2}}{8\pi^2}\exp\left(-2\int_0^{\infty} 
\frac{\rd x \sinh\pi x/6\sinh\pi x/3}{x\sinh\pi x\cosh\pi x/2}\right)Z
\approx 0.0533 Z\ .
\ee

Similarly (\ref{cos}) has a pole at $\theta_{12} = -\ri\pi/3$ 
corresponding to the second breather. The absolute square $|F_2|^2$
of the breather formfactor is found to be
\be
|F_2|^2 = \frac{3^\frac{3}{2}}{8\pi^2}\exp\left(-4\int_0^{\infty} 
\frac{\rd x \cosh\pi x/6\sinh^2\pi x/3}{x\sinh\pi x\cosh\pi x/2}\right)Z
\approx 0.0262 Z\ .
\ee

\section{Dynamical Susceptibilities for a Single Alternating Chain}
\label{sec:1d}
The expression for the imaginary part of the dynamical staggered
susceptibility $\chi'' (\omega,q)$ at $s^2 = \omega^2 -
\frac{v^2}{a_0^2}q^2 < (1 + \sqrt 3)^2M^2$ is given by 
\bea
\Im m\chi''(\omega,q)= 2\pi|F_1|^2\delta(s^2 - M^2) + 2\Re\re 
\frac{|F^{sin}[\theta(s)]_{+-}|^2}{s\sqrt{s^2 - 4M^2}},
\label{nearpi}
\eea
where $\theta(s) = 2\ln(s/2M + \sqrt{s^2/4M^2 - 1})$. Note that all
other formfactors do not contribute to this expression in the
specified range of $s$ as their thresholds are above $(1+\sqrt 3)M$.
Also the normalization $Z$ enters \r{nearpi} only as an overall
factor. 
Since the function $\zeta(\theta)$ vanishes at $\theta = 0$, the
entire formfactor is also finite. Thus the two-particle contribution
to $\chi''(\omega,q)$ exhibits a square-root singularity at the
threshold as a function of $s$.

\begin{figure}[ht]
\noindent
\epsfxsize=0.45\textwidth
(a)
\epsfbox{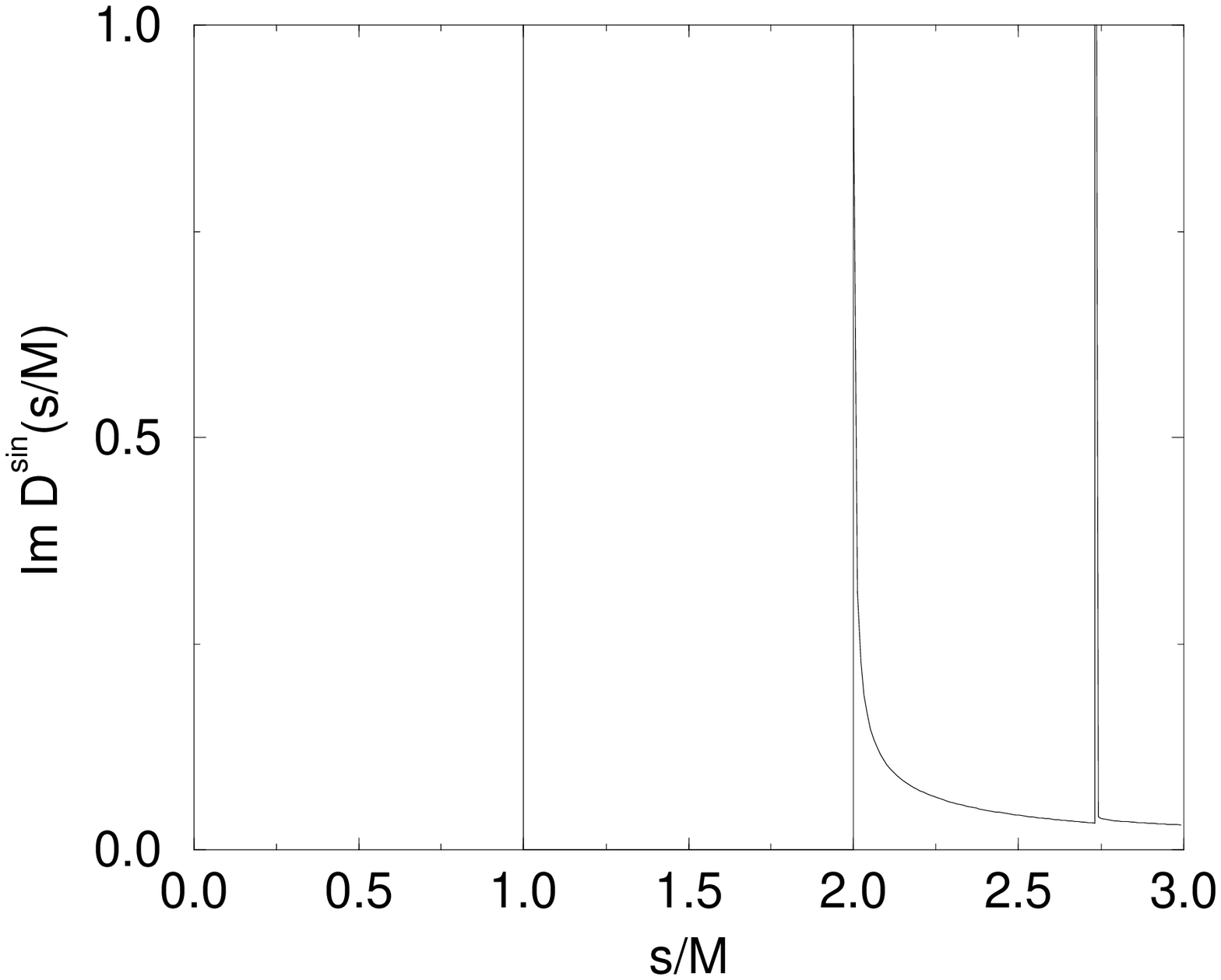}
\epsfxsize=0.45\textwidth
\hfill
(b)
\epsfbox{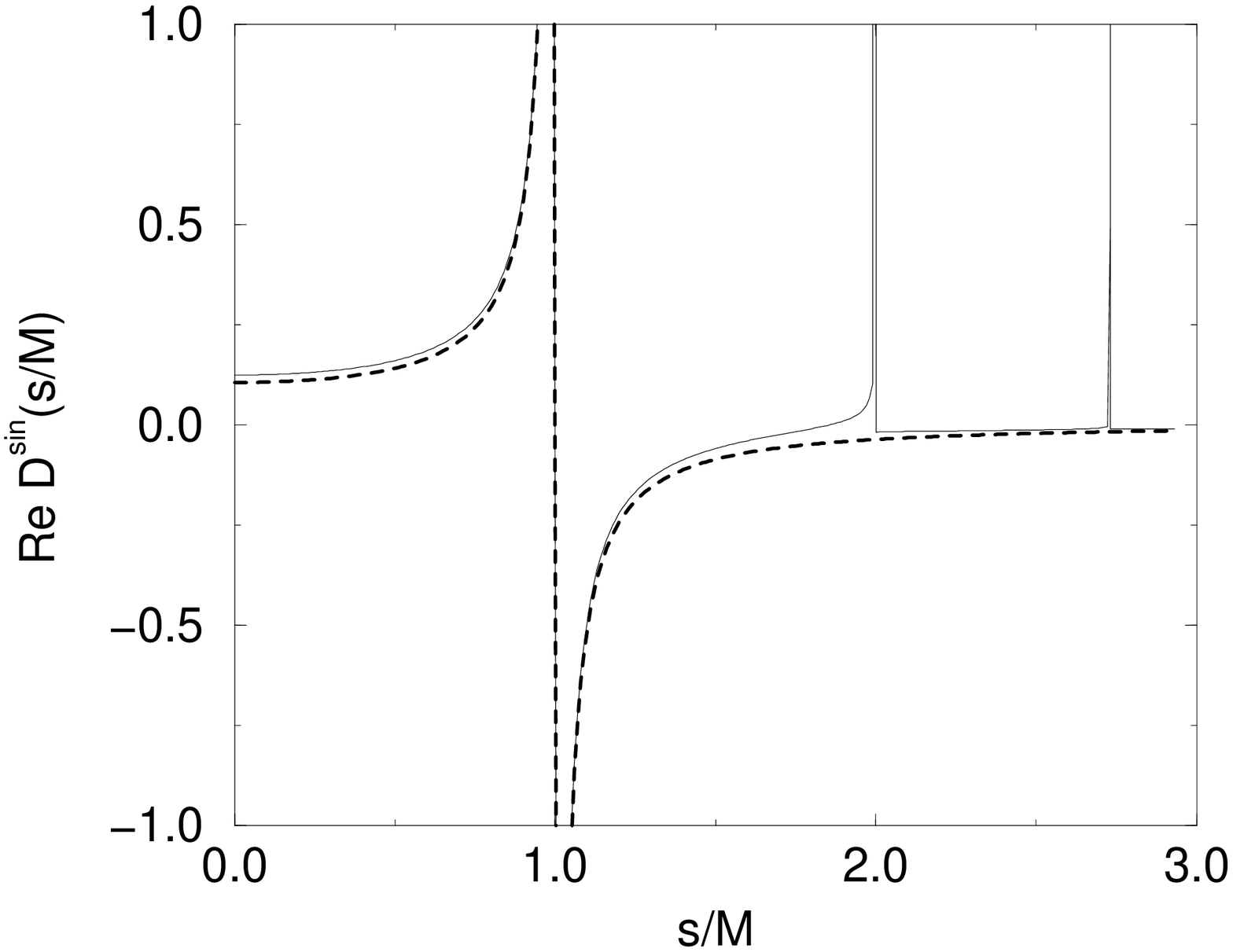}
\caption{\label{fig:chipp}%
Imaginary and real parts (in units of $\frac{Z}{M^2}$) of the
dynamical staggered susceptibility as functions of
$s=\sqrt{\omega^2-\frac{v^2}{a_0^2} q^2}$ for $q\approx\pi$. 
The dashed line depicts the single-mode approximation that takes into
account only the first breather.}
\end{figure}

The breather and $s\bar{s}$ contributions to the real part are found
to be
\be
\Re\re\chi''(\omega,q)= -\Re\re\frac{2 |F_1|^2}{s^2-M^2+i\eps} -
2\int_0^\infty \frac{\rd\theta}{\pi} \frac{s^2-4M^2\cosh^2\frac{\theta}{2}}
{(s^2-4M^2\cosh^2\frac{\theta}{2})^2+\eps^2}
|F^{sin}(\theta)_{+-}|^2\ ,
\label{reals}
\ee
where the factor of $2$ stems from the sum over $+$ and $-$. 
In Fig.~\ref{fig:chipp} we plot both the imaginary and real parts of
$\chi^{\prime\prime}$ 

It is straightforward to repeat the above analysis for the current
operator $\partial_x\Phi$ using the explicit expressions for the
formfactors given in \cite{smirnov}. The contribution of the first
breather leads to to \r{small} with some normalization factor. As there
is very little spectral weight at $q\approx 0$ we concentrate on
$q\approx\pi$ and do not repeat the above analysis for the current
operator. 

For practical purposes it is convenient to have an expression 
interpolating between the small $q$ (\ref{small}) and $q\approx\pi$
(\ref{nearpi}) behaviour. Such an expression giving the dynamical spin
susceptibility in the entire range of $q$ at frequencies below the
continuum may look like
\be
\chi(\omega, q) = \frac{g(q)\sin^2(q/2)}{\omega^2 -
\frac{v^2}{a_0^2}\sin^2q - M^2}\ +\ldots\ ,
\ee
where $g(q)$ is a smooth function interpolating between the
normalizations at $q=0$ and $q=\pi$. The mode $\omega =
\sqrt{\frac{v^2}{a_0^2}\sin^2q + M^2}$ is separated from the particle
continuum by the gap of order of $M$.

Let us now turn to the two-point correlator of cosines. The
contributions of the second breather and the soliton-antisoliton
continuum are given by 
\bea
\Im m D^{cos}(\omega, q) = 2\pi|F_2|^2\delta(s^2 - 3M^2) +2\Re
\re\frac{|F^{\cos}[\theta(s)]_{+-}|^2}{s\sqrt{s^2 - 4M^2}}\ ,
\label{secbr} 
\eea
where the ratio of the single particle residues is universal:
\be
\gamma=\frac{|F_2|^2}{|F_1|^2} = \exp\left( - \int_0^{\infty}\rd x
\frac{\sinh x/3}{x\cosh^2 \frac{x}{2}}\right) \approx 0.49131 \ .
\label{univ}
\ee
Note that the threshold of the breather-breather continuum is also at
$s=2M$. The corresponding contribution is taken into account
in Appendix {\bf A}. The analogous contributions to the real part of
$D^{cos}$ are given by 
\be
\Re\re D^{cos}(\omega,q)= -\Re\re\frac{2 |F_2|^2}{s^2-3M^2+i\eps}
-2\int_0^\infty \frac{\rd\theta}{\pi}\frac{s^2-4M^2\cosh^2\frac{\theta}{2}}
{(s^2-4M^2\cosh^2\frac{\theta}{2})^2+\eps^2}|F^{cos}(\theta)_{+-}|^2\ .
\label{realc}
\ee
The remaining integrals in \r{nearpi},\r{reals},\r{secbr} and
\r{realc} have to be calculated numerically. We find that at small $s$
the contributions of the two-particle continua to the real parts of
both correlators are of the same magnitude as the single-particle
contributions from the breather states. As far as a single chain is
concerned a single-mode approximation taking into account only the
one-particle states is therefore very poor at small $s$.
\begin{figure}[ht]
\noindent
\epsfxsize=0.45\textwidth
(a)
\epsfbox{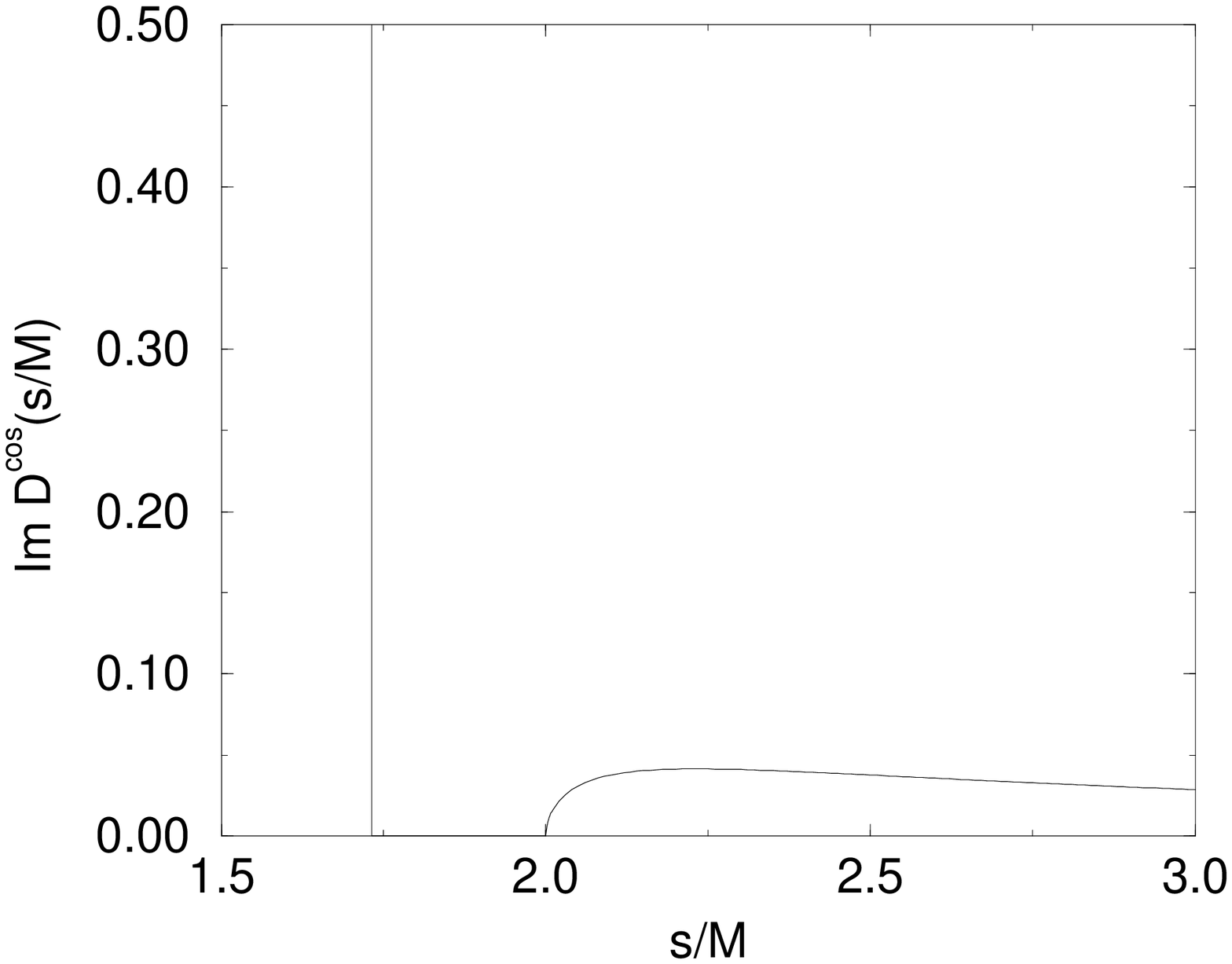}
\epsfxsize=0.45\textwidth
\hfill
(b)
\epsfbox{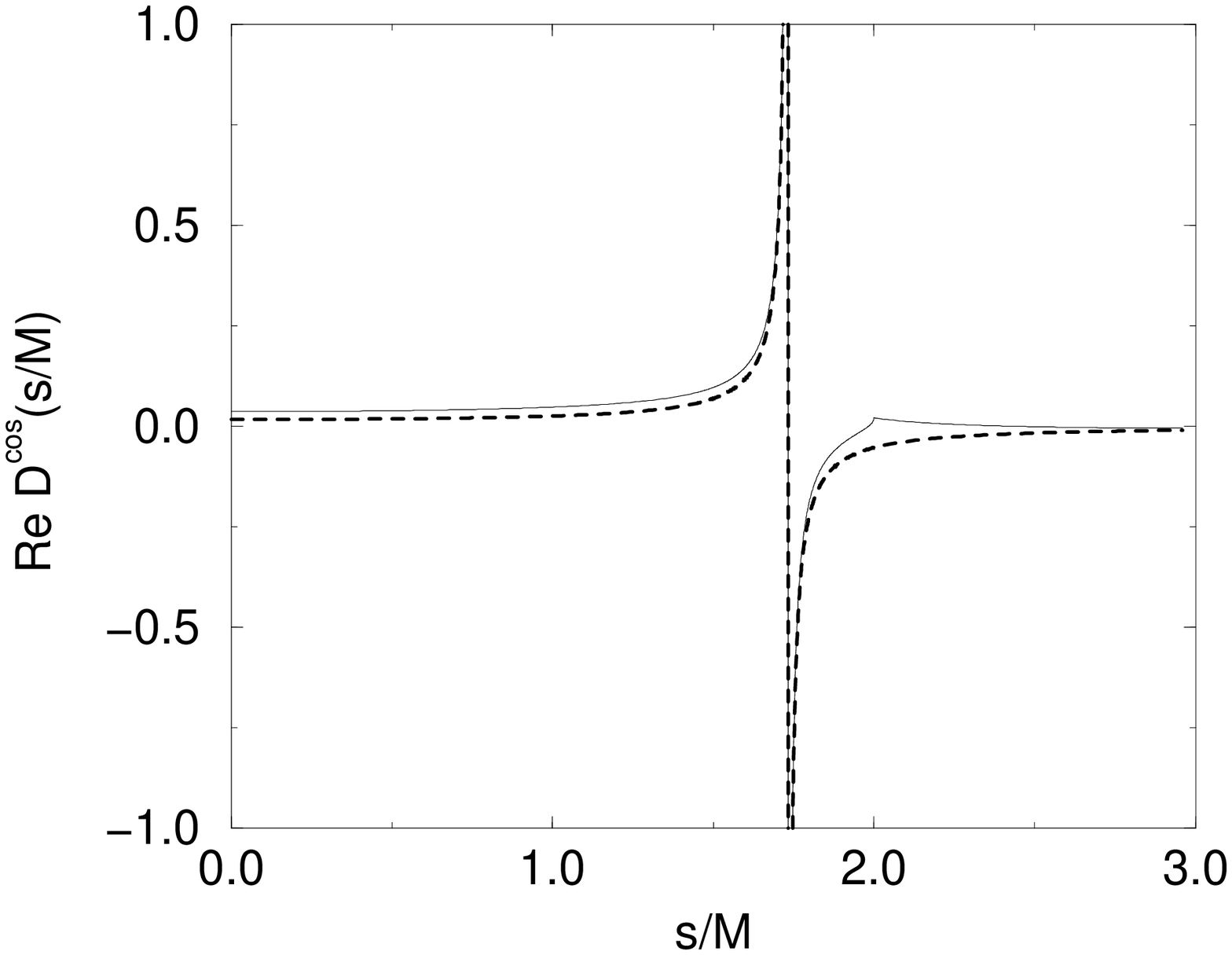}
\caption{\label{fig:cos}%
Imaginary and real parts of $D^{cos}$ (in units of $\frac{Z}{M^2}$) as
functions of $s=\sqrt{\omega^2-\frac{v^2}{a_0^2} q^2}$ for
$q\approx\pi$.
}
\end{figure}

\section{RPA analysis of the interchain interactions}

Let us now take into account the interchain interactions (both of
spin and staggered energy densities) in (\ref{H1}). This is
accomplished through an RPA analysis along the lines of \cite{wen}. 
RPA becomes exact in the limit of an infinite number of neighbouring
chains. 

In the RPA we obtain the following expression for the correlation
function of energy densities
\be
\chi_\eps(s,\vec{k})=\la\!\la\epsilon(- \omega, -q; -
\vec{k})\epsilon(\omega, q; \vec{k}) \ra\!\ra = \frac{D^{cos}(s)}{1 -
D^{cos}(s)\ J(\vec{k})} 
\label{rpa}
\ee
where $J(\vec{k})=2[|J_a|\cos(k_x)+|J_b|\cos(k_y)]$. 
Similarly the dynamical staggered susceptibility is given by
\be
\chi^{zz}(s,\vec{k})=\la\!\la S^z(- \omega, -q; -
\vec{k})S^z(\omega, q; \vec{k} )\ra\!\ra = \frac{D^{sin}(s)}{1 +
D^{sin}(s)\ K(\vec{k})}\ , 
\label{rpaspin}
\ee
where $K(\vec{k})=2[K_a\cos(k_x)+K_b\cos(k_y)]$. 
Note that in the present approximation $\chi_\eps(s,\vec{k})$ is only
affected by the interchain interactions of staggered energy densities
whereas $\chi^{zz}(s,\vec{k})$ only ``sees'' the interchain
interactions of spin densities. The reason for this decoupling is that
in the sine-Gordon theory describing the individual chains
\be
\langle \eps(t,x) S^z(0,0)\rangle =0\ ,
\ee
because $\eps(t,x)$ is even under charge conjugation whereas
$S^z(0,0)$ is odd. By rotational invariance this implies
$\langle \eps(t,x) \vec{S}(0,0)\rangle =0$. Note that the RPA is
particularly simple as we have taken into account only interchain
interactions of the staggered part of the spin density and neglected
the smooth part as being less relevant. If we take these
subleading terms into account the RPA acquires a matrix structure like
in \cite{schulz} as the sectors $q\approx\pi$ and $q\approx 0$ become
coupled. An RPA analysis then requires the calculation of formfactors
of the current operator. We will discuss this refined RPA in a
separate publication \cite{unpub}. 

The response functions $\chi^{\alpha\alpha}(s,\vec{k})$ and
$\chi_\eps(s,\vec{k})$ can be easily calculated numerically for given
values of $J_{a,b}$ and $K_{a,b}$ by using the expressions for
$D^{sin}(s)$ and $D^{cos}(s)$ obtained in section \ref{sec:1d}.
We note that the pole in the dynamical magnetic susceptibilities 
$\chi^{\alpha\alpha}(s,\vec{k})$ corresponds to the light breather
which has quantum number $S^z=1$, whereas the pole in 
$\chi_\eps(s,\vec{k})$ is due to the heavy breather with $S^z=0$. 
The magnetic mode can be measured directly by neutron scattering
whereas the $S^z=0$ excitation can be probed by measuring the phonon
spectrum which will exhibit a softening.

In order to visualize our results we now plot them for a particular
choice of parameters. Being aware that our theory probably cannot be
applied to ${\rm CuGeO_3}$ where $\alpha >\alpha_c$ we nevertheless
choose these parameters to reproduce the dispersions of magnetic
excitations in that material. We take $\kappa\approx 0.8$, $M\approx
4.58 meV$ and $\frac{|K_b|Z}{M^2}\approx 3.08$,
$\frac{|K_a|Z}{M^2}\approx 0.25$.
The value of $\kappa$ is chosen such that the dispersion in
$z$-direction reproduces the experimental fit of \cite{reg}, whereas
the other conditions follow from the experimental band-gaps for
$\vec{k}$-vectors $(0,1,\frac{1}{2})$ ($\approx 2meV$),
$(0,0,\frac{1}{2})$ ($\approx 5.7 meV$) and
$(\frac{1}{2},1,\frac{1}{2})$ ($\approx 2.6 meV$). 

\begin{figure}[ht]
\noindent
\epsfxsize=0.45\textwidth
(a)
\epsfbox{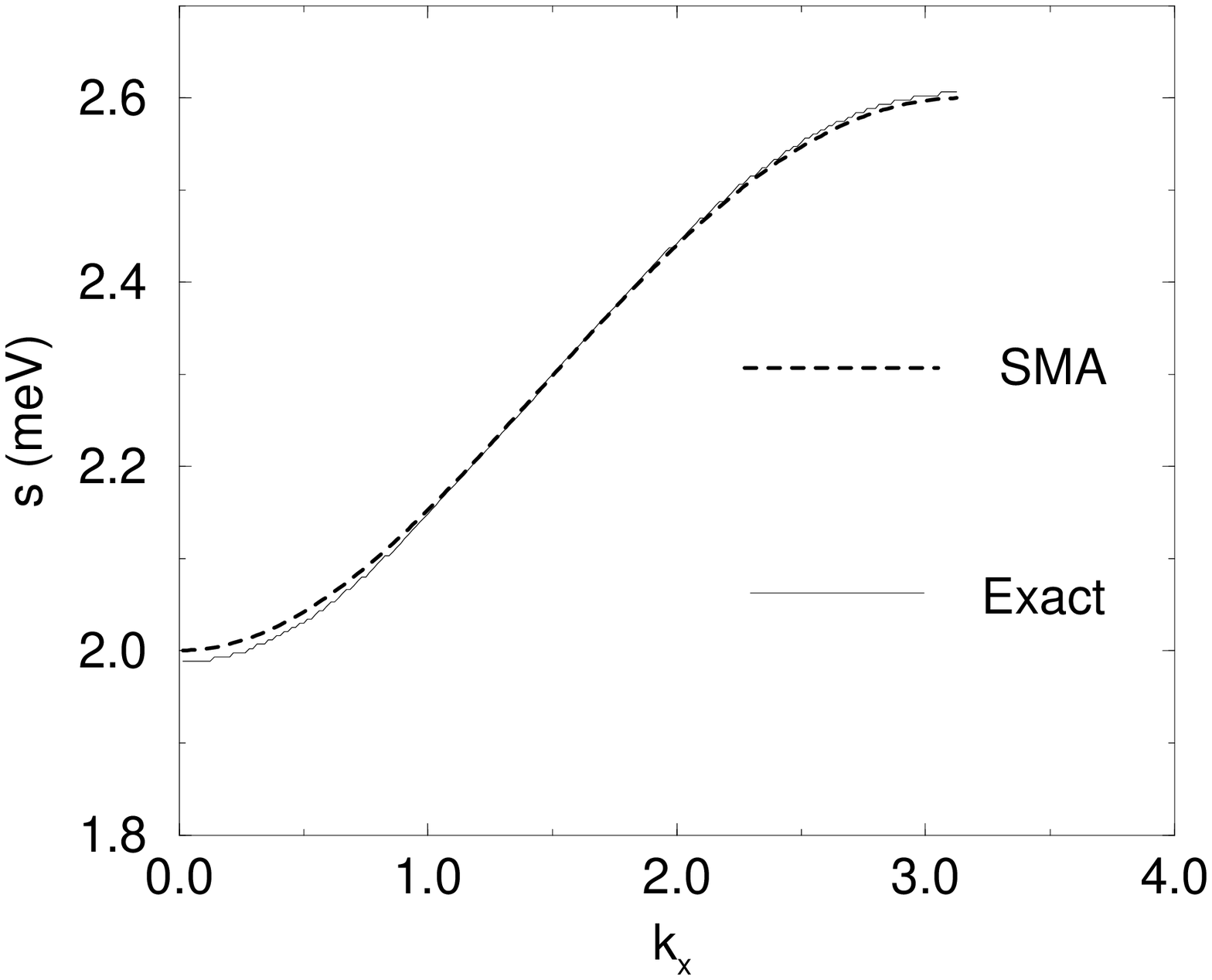}
\epsfxsize=0.45\textwidth
(b)
\epsfbox{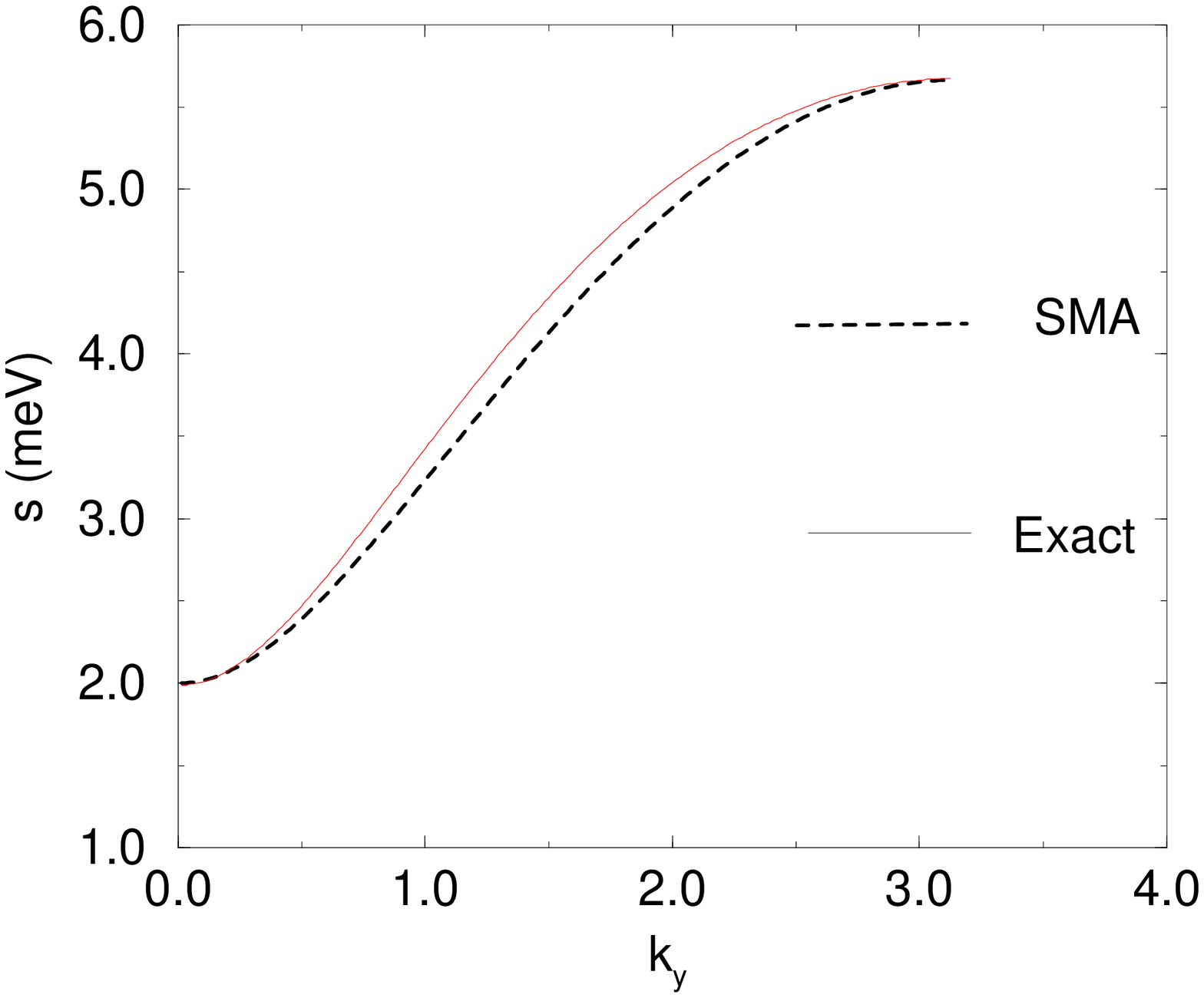}
\caption{\label{fig:dispy_sP}%
Spin-wave dispersions in $x$ (a) and $y$ (b) directions.
}
\end{figure}

\begin{figure}[ht]
\noindent
\epsfxsize=0.45\textwidth
(a)
\epsfbox{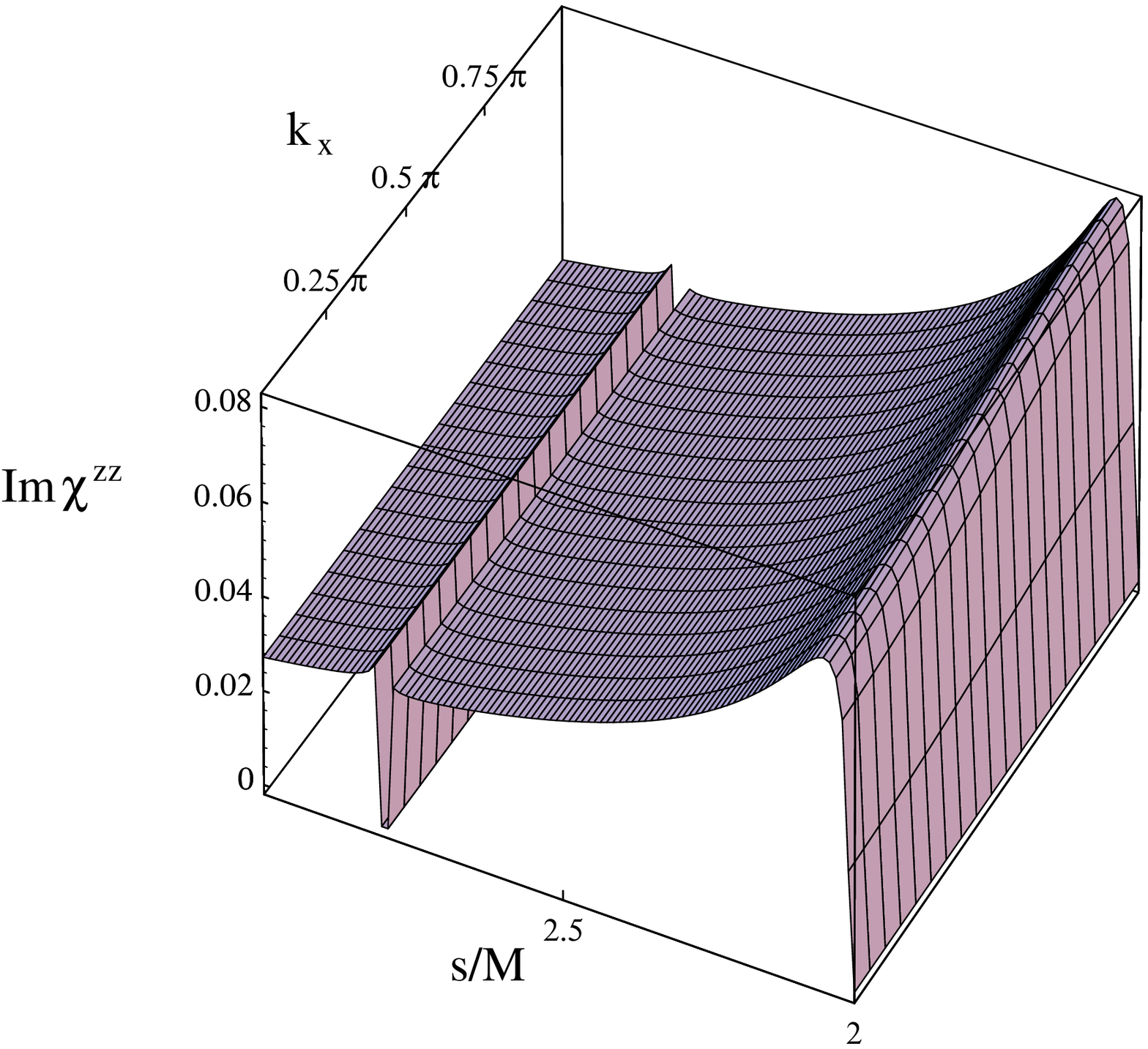}
\epsfxsize=0.45\textwidth
(b)
\epsfbox{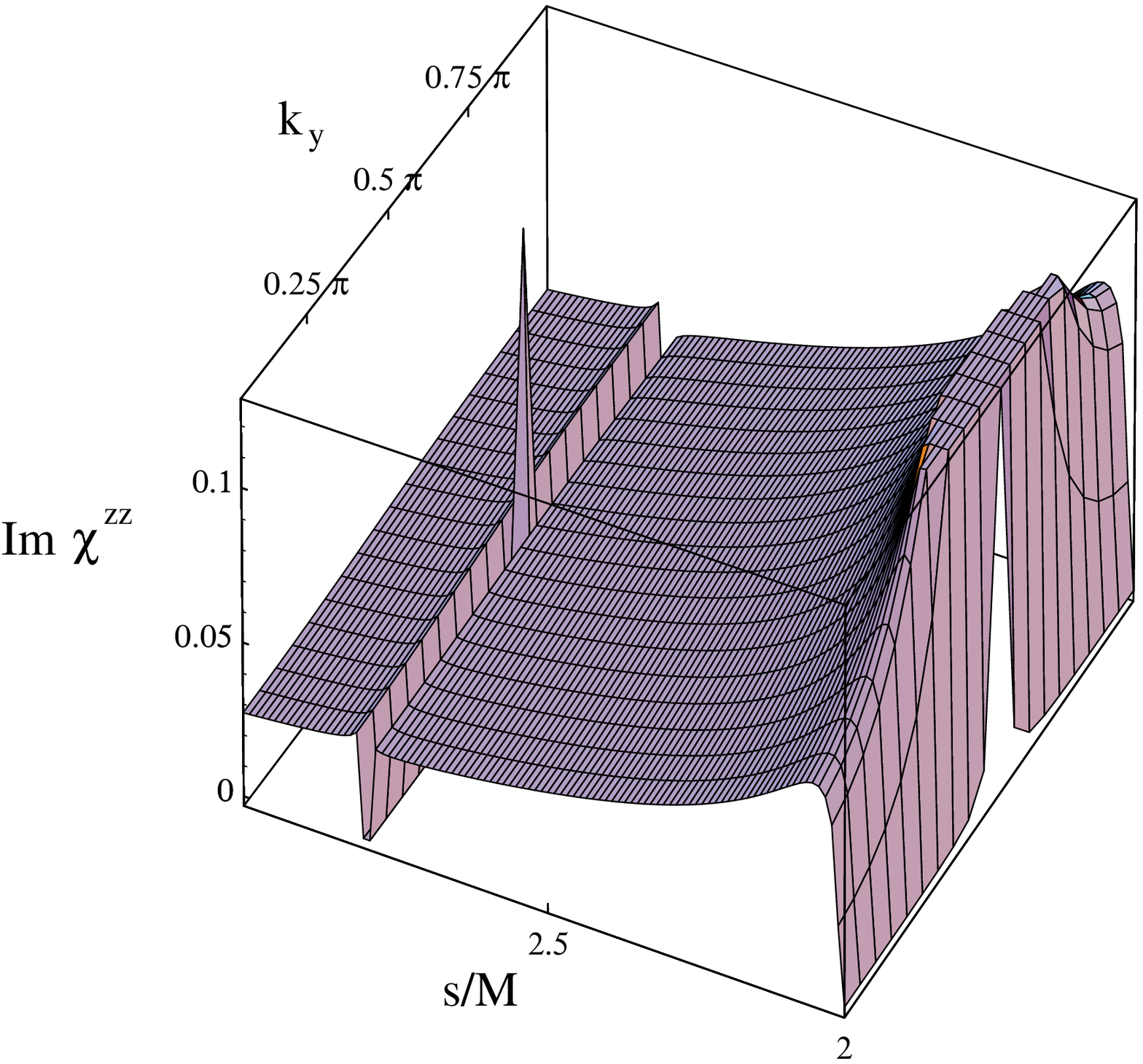}
\caption{\label{fig:imchix_sP}%
Imaginary part (in arbitrary units) of $\chi^{zz}(\omega,q;\vec{k})$
in $x$ (a) and $y$ (b) directions. 
}
\end{figure}

Because of Lorentz-invariance the energy $\omega$ and the
z-component of the momentum $q$ only enter in the combination
$s=\sqrt{\omega^2-\frac{v^2}{a_0^2}q^2}$.
In Fig.~\ref{fig:dispy_sP} (a) we plot the spin-wave dispersion in 
$x$-direction ($k_y=0$, $k_x\in (0,\pi)$) and in 
Fig.~\ref{fig:dispy_sP} (b) in $x$-direction ($k_x=0$, $k_y\in
(0,\pi)$).
We see that the single-mode approximation (SMA) in which all
multiparticle contributions to the dynamical susceptibilities are
neglected gives essentially the same result as the exact
treatment. We note that (by construction) the fits \cite{reg} to the
experimental results are essentially identical to the SMA as far as
dispersion relations are concerned.
Let us now turn to the multiparticle continuum. The imaginary part of
the dynamical staggered susceptibility is directly measurable by
Neutron scattering. The position of its poles yields the dispersion
discussed above. The incoherent part (as a function of $s$ and $k_{x,y}$)
is plotted in Fig.~\ref{fig:imchix_sP} (a) and (b) respectively.

In the RPA the 2-particle continuum starts at $s=2M\approx 9meV$. This
is in disagreement with experiment for ${\rm CuGeO_3}$. 
This may be because $\alpha >\alpha_c$ or because the dispersion
in $b$-direction is rather strong, which in turn forces $M$ to be
large in order to get a reasonable fit to the experimental data. 

\section{Antiferromagnetic Order}

In the continuum limit the model \r{stchain} is also equivalent to
the sine-Gordon model with $\beta^2 = 2\pi$ \cite{schulz}. The
bosonization formulas are now
\bea
\vec{\bf S} (x) &=& \vec{\bf J}_R (x) + \vec{\bf J}_L(x) + (-1)^n
\vec{\bf n} (x)\ ,\nn 
J^z_{R,L}&=&\frac{1}{2\sqrt{2\pi}}\left(\partial_x\Phi\mp\Pi\right)\
,\quad J^+_{R,L}=\frac{1}{2\pi a_0} \exp\left(\mp
i\sqrt{2\pi}(\Phi\mp\Theta)\right),\nn 
n^z (x) &=& - \frac{\lambda}{\pi a_0} \cos \sqrt{2 \pi} \Phi (x), \quad
n^\pm (x) = \frac{\lambda}{\pi a_0} \exp [ \pm \ri \sqrt{2 \pi}
\Theta (x)].
\label{boso2}
\eea
Since the staggered magnetization is proportional to
$\sin(\sqrt{2\pi}\Phi)$ the bosonized single-chain Hamiltonian is
given by 
\be
H=\frac{v}{2}\int \rd x\left[\pi^2(x)+(\partial_x\Phi(x))^2\right]
-\frac{h\lambda}{\pi a_0}\int \rd x\ \sin\sqrt{2\pi}\Phi
- 2N J_\perp m_0^2\ . 
\ee
Following through the same steps as in the spin-Peierls case we find
the following mean-field results for staggered magnetization and mass
gap 
\bea
m_0={\cal C}\left|\frac{J_\perp}{J}\right|^\frac{1}{2}\ ,\qquad
M={\cal C}^\prime|J_\perp|\ ,
\eea
where the ratio of ${\cal C}$ and ${\cal C}^\prime$ is given by
(\ref{ratio}). 
This relation can be used to determine the transverse coupling
$J_\perp$ in terms of $J$ and the directly measurable quantities $M$
and $m_0$ as follows. The gap $M$ is equal to $\omega(\pi,0,\pi)$
which is found experimentally to be $11.0\pm 0.5 meV$ in ${\rm
KCuF_3}$ \cite{satija}. The average magnetic moment is $m_0\approx
0.27$ \cite{welz} and $J\approx 53.17\pm 0.25 meV$\cite{tenn}. Using
these values we find
\be
|J_\perp|=\frac{1}{J}\left(\frac{{\cal C}M}{{\cal
C}^\prime m_0}\right)^2\approx 0.96 meV\ .
\ee
Let us now turn to the correlation functions for a single chain.
The correlator $\la\!\la S^z S^z\ra\!\ra$ at small $q$ is
still given by Eq.(\ref{small}), but for $q \approx \pi$ is given
by Eq.(\ref{secbr}). Therefore around $q = \pi$ the pole is at $s
= \sqrt 3M$
\be
\chi^{zz}(\omega, q) \equiv D^{cos}(s)= \frac{2F_2^2}{3M^2 +
\frac{v^2}{a_0^2}(\pi - q)^2 - \omega^2} + \mbox{incoherent}\ ,
\label{afmczz}
\ee
where ``incoherent'' denotes multiparticle contributions. Some of
these are determined exactly in Appendix A. A plot of $D^{cos}(s)$
is shown in Fig.\ref{fig:cos}.

Correlation functions of transverse components of the staggered
magnetization are given by Eq.(\ref{nearpi}) and have a pole at $s =
M$ 
\be
\chi^{xx}(\omega, q)\equiv D^{sin}(s) = \frac{2F_1^2}{M^2 +
\frac{v^2}{a_0^2}(\pi - q)^2 - \omega^2} + \mbox{incoherent}\ .
\label{afmcp}
\ee
Here we have used the fact that in the continuum model the correlation
functions of $\cos\sqrt{2\pi}\theta$, $\sin\sqrt{2\pi}\theta$ and
$\sin\sqrt{2\pi}\Phi$ are equal due to the SU(2) symmetry present at
$\beta=\sqrt{2\pi}$. A plot of $D^{sin}(s)$ is shown in
Fig.~\ref{fig:chipp}. 

The difference in the correlation functions (\ref{afmczz}) and
(\ref{afmcp}) is obviously related to the broken rotational symmetry
of the Hamiltonian (\ref{stchain}).
Next we take into account the interchain interaction by an RPA
analysis. This yields the following expression for the longitudinal
dynamical susceptibility 
\bea
\chi^{zz}(\omega, q, \vec{k}) &=&
\frac{D^{cos}(s)}{1-2|J_{\perp}|(\cos k_x+\cos k_y)D^{cos}(s)}\nn 
\chi^{xx}(\omega, q,\vec{k})&=& \frac{D^{sin}(s)}{1- 2|J_\perp|
(\cos k_x +\cos k_y)D^{sin}(s)} \ .
\eea
Here we again have taken into account only the staggered part of the
spin density as it gives the most relevant contribution to the
interchain interaction. As a result the RPA expression for the
susceptibilities are of scalar rather than matrix form \cite{schulz}.
The transverse susceptibility must have a pole at the Neel wave vector
$(0,0,\pi)$ as the spin $SU(2)$ symmetry is spontaneously broken. This
leads to the requirement that
\be
D^{sin}(0)=\frac{1}{4|J_{\perp}|}\approx 0.12509 \frac{Z}{M^2}\ ,
\ee
which fixes the normalization $Z$ in terms of the transverse coupling
and the breather mass as $Z\approx 1.999\frac{M^2}{|J_\perp|}$.
The normalization of the correlator of cosines then follows to be
\be
D^{cos}(0)\approx\frac{0.07443}{|J_{\perp}|}\ .
\ee

\begin{figure}[ht]
\noindent
\epsfxsize=0.45\textwidth
(a)
\epsfbox{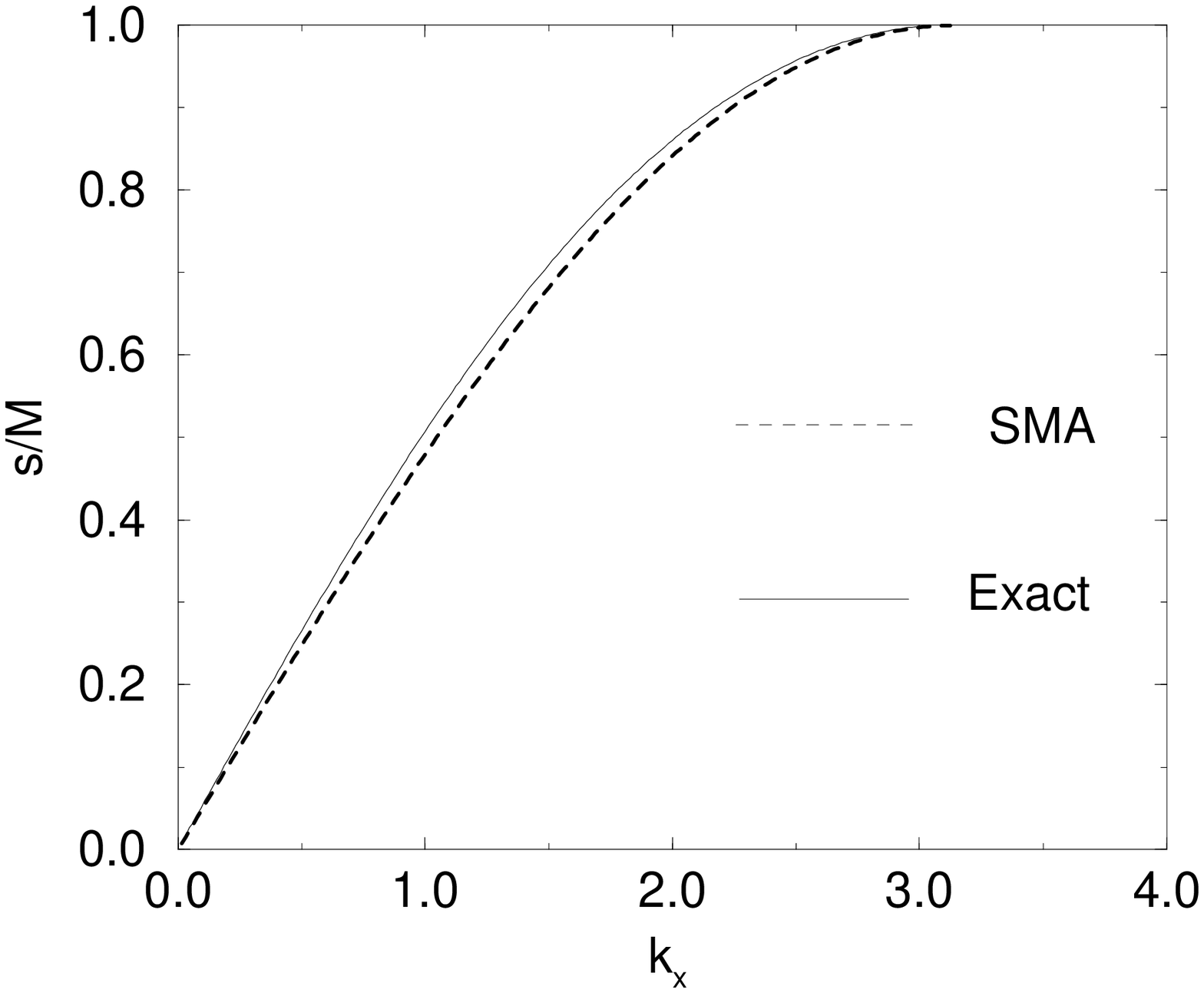}
\epsfxsize=0.45\textwidth
(b)
\epsfbox{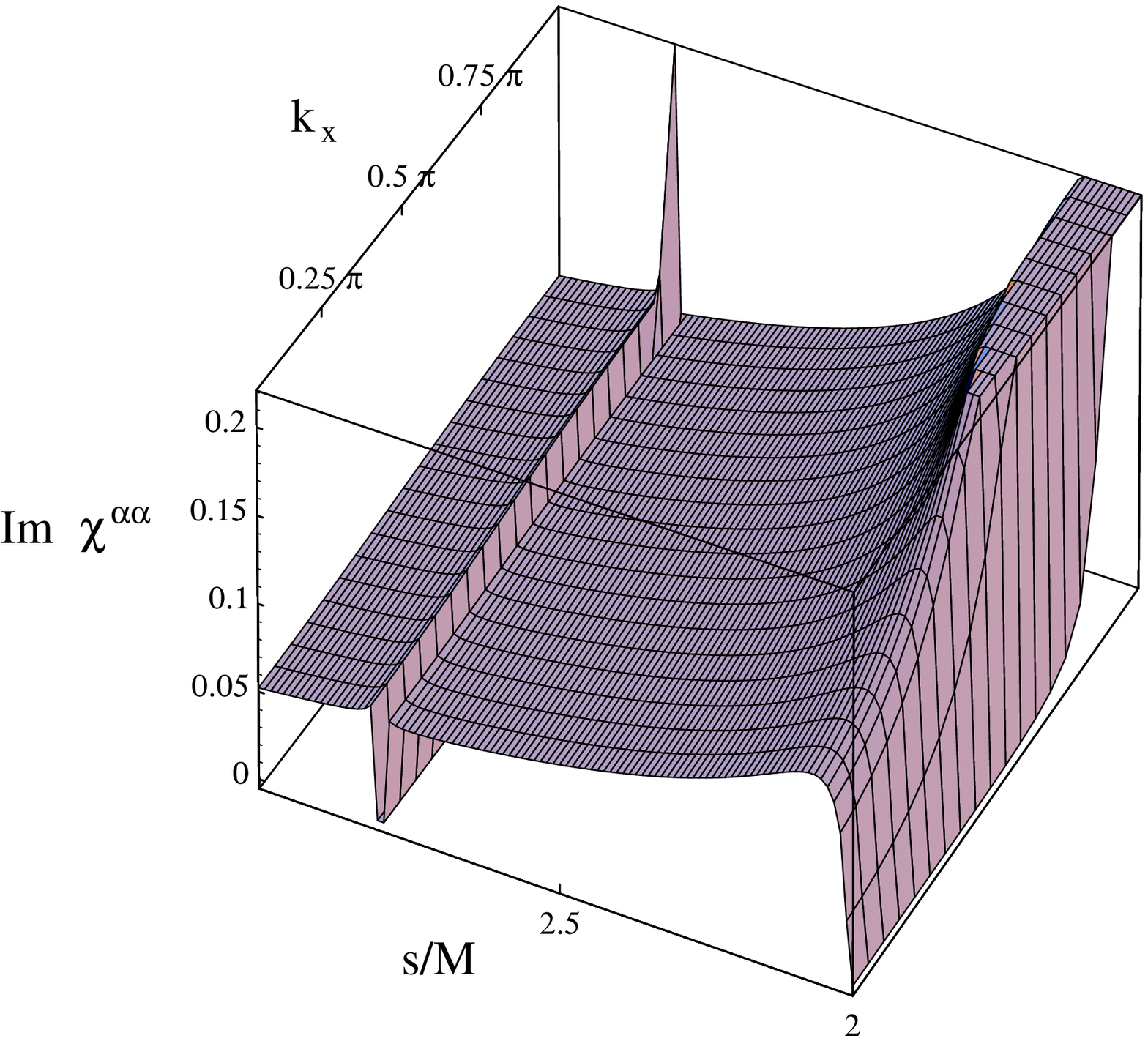}
\caption{\label{fig:chiperp}%
(a) Spin-wave dispersion as a function of $k_x$ for $k_y=0$.
(b) Imaginary part (in arbitrary units) of $\chi^{\perp}$ for $s> 2$
as a function of $k_x$ for $k_y=0$. }
\end{figure}

The Goldstone mode associated with the zero energy pole in $\chi^\perp$
is a spin-wave moving in $z$-direction and its dispersion is found
from the singularities of $\chi^{\perp}$. Due to Lorentz invariance
$D^{sin}$ depends only on $s$ rather than on $\omega$ and $q$
independently. This immediately implies that the spin-wave dispersion
for $q\approx \pi$ is
\be
\omega^2(q,\vec{k}) = \frac{v^2}{a_0^2}(\pi-q)^2+M^2(1-\frac{\cos
k_x+\cos k_y}{2})\approx \frac{\pi^2 J^2}{4}(\pi-q)^2
+M^2(1-\frac{\cos k_x+\cos k_y}{2})\ .
\label{swdisp}
\ee
This is in very good agreement with experiment being almost
identical to the fit used in \cite{satija}. In
Fig.~\ref{fig:chiperp} (a) we plot the spin-wave dispersion for
$k_y=0$ and $k_x\in (0,\pi)$.

%
We see that the SMA works extremely well for all values of $k_x$.
The imaginary part of the dynamical susceptibility is directly
measurable by neutron scattering. We find 
\be
\Im m\chi^{\perp}(\omega,q,\vec{k}) = \frac{\pi}{2|J_\perp|}
\delta(\frac{s^2}{M^2}-1+\frac{\cos k_x+\cos k_y}{2})+{\rm
incoherent}\ , 
\ee
where we have used the SMA to get the delta-function part. The
incoherent part is plotted in Fig.~\ref{fig:chiperp} (b).
%
We see that there is in general no singularity at the threshold of the
light breather-heavy breather continuum except at $k_x\to\pi$ where 
$\Im m\chi^{\perp}(\omega,q,\vec{k}) \equiv \Im m D^{sin}(\omega,q)$
so that we recover the pure 1-D result.
The situation for the soliton-antisoliton continuum is analogous.

\begin{figure}[ht]
\noindent
\epsfxsize=0.45\textwidth
\epsfbox{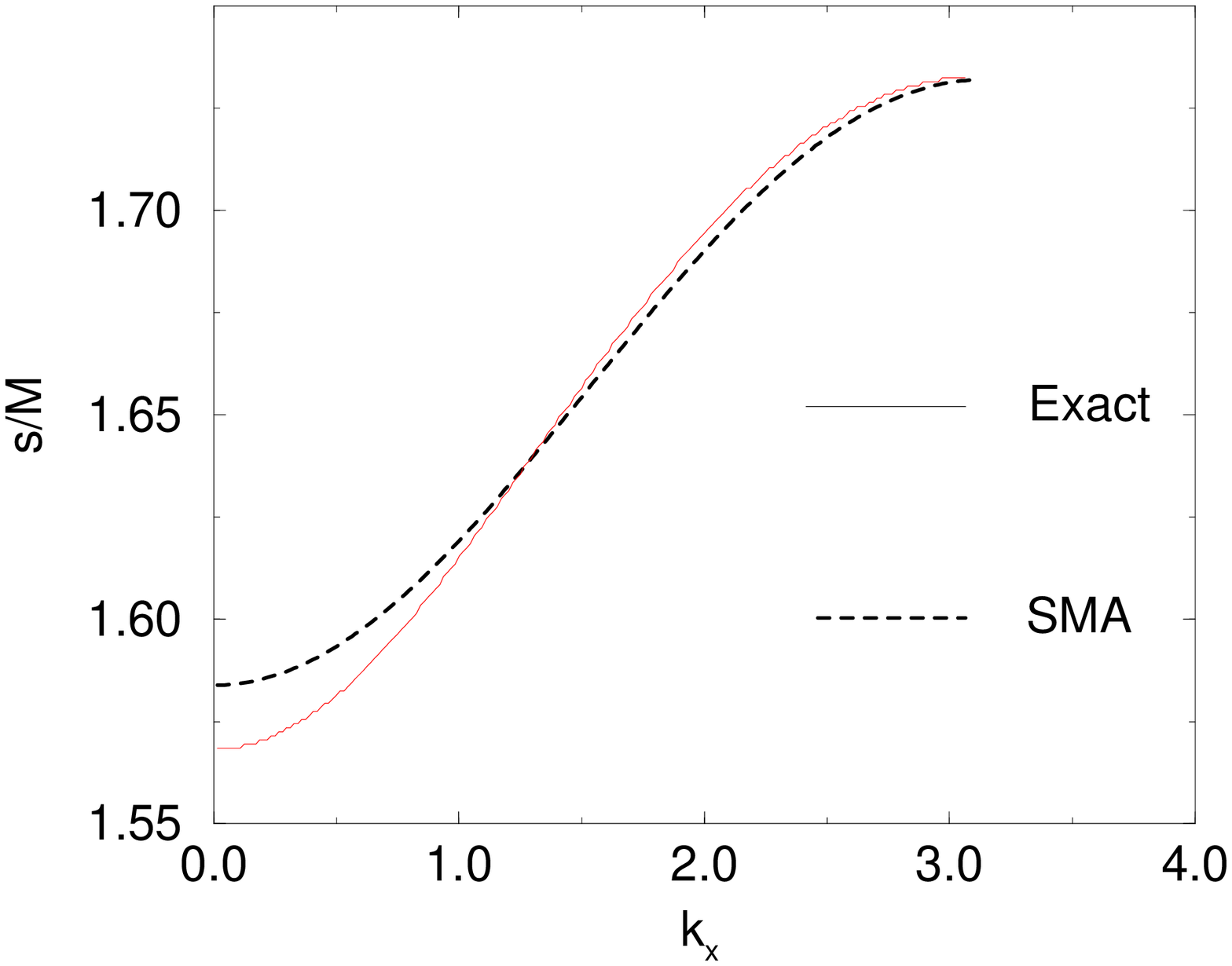}
\epsfxsize=0.45\textwidth
\epsfbox{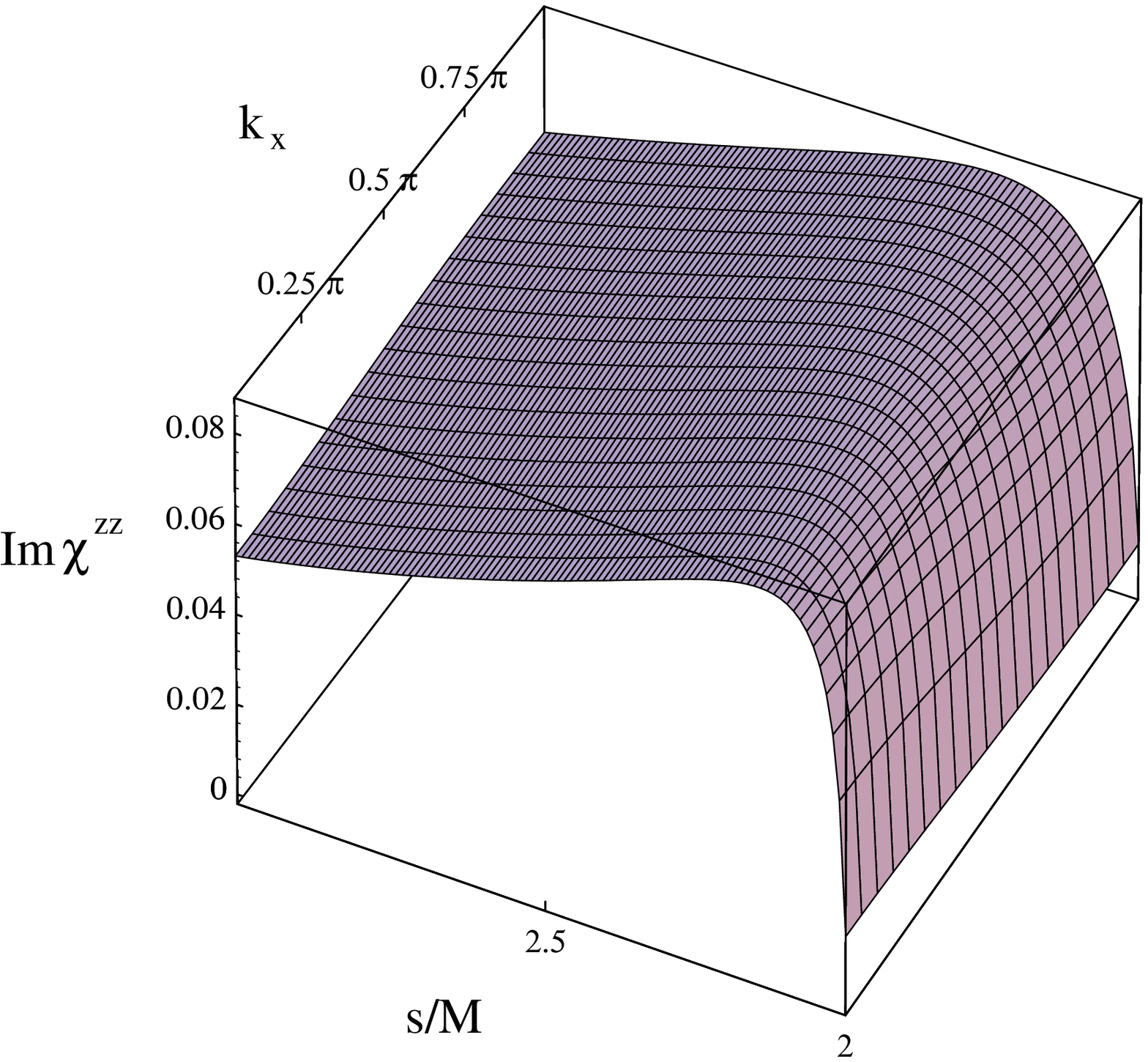}
\caption{\label{fig:dispx}%
(a) Dispersion of the longitudinal mode as a function of $k_x$ for
$k_y=0$.
(b) Imaginary part (in arbitrary units) of $\chi^{zz}$ for $s\geq 2$
as a function of $k_x$ for $k_y=0$. }
\end{figure}

Let us now turn to the longitudial susceptibility. In the SMA there is
a pole in $\chi^{zz}$ at  
\be
\omega^2(q,\vec{k})=\frac{v^2}{a_0^2}(\pi-q)^2
+M^2[3-\frac{\gamma}{2}(\cos k_x+\cos k_y)]\ ,  
\label{disp}
\ee
where $\gamma$ is given by \r{univ}. This is compared to the exact
result for the case where $k_y=0$ in Fig.~\ref{fig:dispx} (a). We see
that the corrections to the SMA result are very small.

Using the SMA (which we know from Fig.~\ref{fig:dispx}(a) to be an
excellent approximation) to extract the coherent delta-function part
we find 
\be
\Im m\chi^{zz}(\omega,q,\vec{k}) = \frac{\pi\gamma}{4|J_\perp|}
\delta(\frac{s^2}{M^2}-3+\frac{\gamma}{2}(\cos k_x+\cos k_y))+{\rm
incoherent}\ .
\ee
The incoherent part is plotted in Fig.~\ref{fig:dispx}(b).

\begin{center}
{\bf Acknowledgements}
\end{center}
We thank R. Cowley, S. Nagler, L.- P. Regnault and F. Smirnov for
interesting and valuable conversations. 

\vskip .5cm
\appendix{\centerline{\bf APPENDIX A: MULTIPARTICLE FORMFACTORS}}
\vskip .5cm
In this appendix we consider multiparticle formfactors. We start with
2-soliton 2-antisoliton formfactors and explicitly derive the
related three and two-particle formfactors. The extension to 
$n$-soliton $n$-antisoliton formfactors ($n=3,4,\ldots$) is
straightforward and will not be discussed here. The formfactor
expansion for 2-point correlation functions is found to be rapidly
converging and for small $s$ it is essentially sufficient to take into
account 2-particle formfactors only. Note that most
of the  formulas below are to be understood in terms of analytic
continuation of $\zeta(\theta)$. A useful formula is
\bea
|\zeta(\theta+i\a)\zeta(\theta-i\a)|^2&=&\frac{c^4}{4}(\cosh\theta-\cos\a)^2
\exp\left\{4\int_0^{\infty}\rd x \frac{\cosh x/6}{x\sinh x\cosh
\frac{x}{2}}(1-\cos\frac{x\theta}{\pi}\cosh\frac{\pi-\a}{\pi}x)\right\}
\nn &\times& 
\frac{\sinh^2\theta\cos^2\a+(\cosh\theta\sin\a-\sin\frac{\pi}{3})^2}
{\sinh^2\theta\cos^2\a+(\cosh\theta\sin\a+\sin\frac{\pi}{3})^2}\ .
\eea

Multiparticle formfactors of (quasi)local operators were studied in
detail by Smirnov \cite{smirnov}. From his work the 2-soliton
2-antisoliton formfactor for $\cos\sqrt{2\pi}\Phi$ is
straightforwardly extracted
\bea
&&F^{cos}(\theta_1,\theta_2,\theta_3,\theta_4)_{--++} =
\la 0|\cos\sqrt{2\pi}\Phi|\theta_1,\theta_2,\theta_3,\theta_4
\ra_{--++} = \nn
&&\hskip50pt 2\pi (2d)^2 \sqrt{Z}\ 3\left(\sum_{l=1}^4 e^{\theta_l}\right)
\left(\sum_{m=1}^4 e^{-\theta_m}\right)
\sum_{i<j}e^{\theta_i+\theta_j}\prod_{i<k}\zeta(\theta_i-\theta_k)
e^{-\frac{1}{2}\sum_j\theta_j}\ \times \nn
&&\hskip50pt \times\quad
\cosh\frac{3}{2}(\theta_3+\theta_4-\theta_1-\theta_2) 
\prod_{i=1}^2\prod_{j=3}^4\frac{1}{\sinh 3(\theta_j-\theta_i)}\ .
\label{cos4}
\eea
Orderings other than $--++$ are obtained from \r{cos4} by using the
generalization of \r{24} {\sl e.g.}
\be
F^{cos}(\theta_1,\theta_2,\theta_3,\theta_4)_{-+-+} =
S_0(\theta_3-\theta_2)\
F^{cos}(\theta_1,\theta_3,\theta_2,\theta_4)_{--++}\ .
\ee
It is easy to verify that the soliton-antisoliton formfactor of 
$\cos\sqrt{2\pi}\Phi$ is obtained from (\ref{cos4}) {\sl via} the
annihilation pole condition (\ref{annpole}).
The 2-soliton 2-antisoliton formfactor for $\sin\sqrt{2\pi}\Phi$ is
very similar to the one for $\cos\sqrt{2\pi}\Phi$
\be
F^{sin}(\theta_1,\theta_2,\theta_3,\theta_4)_{--++} =
-\ri \tanh\frac{3}{2}(\theta_3+\theta_4-\theta_1-\theta_2)
F^{cos}(\theta_1,\theta_2,\theta_3,\theta_4)_{--++}\ .
\ee
The residue at the annihilation pole (times $i$) now yields the
soliton-antisoliton formfactor of $\sin\sqrt{2\pi}\Phi$.

Breather formfactors are obtained from the residues of \r{cos4} at its
poles. In the soliton-antisoliton-even breather sector we find
\bea
F^{cos}(\theta_1,\theta_2,\theta_3)_{-+2} &=&
-2\pi \frac{(2d)^2 \sqrt{Z}}{2^\frac{3}{2}3^\frac{1}{4}}
\frac{e^{-\frac{1}{2}(\theta_1+\theta_2)-\theta_3}}{\cosh\frac{3\theta_{21}}{2}
\cosh 3\theta_{31} \cosh 3\theta_{32}}\ \zeta(\theta_{12})
\zeta(-\ri\frac{\pi}{3}) \zeta(\theta_{13}+\ri\frac{\pi}{6})\nn
&\times&
\zeta(\theta_{13}-\ri\frac{\pi}{6})
\zeta(\theta_{23}+\ri\frac{\pi}{6})
\zeta(\theta_{23}-\ri\frac{\pi}{6})
[e^{\theta_1}+e^{\theta_2}+\sqrt{3}e^{\theta_3}]\nn
&\times& [e^{-\theta_1}+e^{-\theta_2}+\sqrt{3}e^{-\theta_3}]
[e^{\theta_1+\theta_2}+\sqrt{3}e^{\theta_3}(e^{\theta_1}+e^{\theta_2})+e^{2\theta_3}].
\eea
The corresponding formfactor for $\sin\sqrt{2\pi}\Phi$ is
\be
F^{sin}(\theta_1,\theta_2,\theta_3)_{-+2} =
-\ri \coth\frac{3\theta_{21}}{2}\
F^{cos}(\theta_1,\theta_2,\theta_3)_{-+2}\ ,
\ee
and different orderings are obtained by the appropriate generalization
of \r{24} {\sl e.g.}
\be
F^{cos}(\theta_1,\theta_2,\theta_3)_{2-+} =
F^{cos}(\theta_2,\theta_3,\theta_1)_{-+2}\ S_{1,2}(\theta_{21})\
S_{1,2}(\theta_{31})\ .
\ee
The residue at the annihilation pole (times $i$) in
$F^{cos}(\theta_1,\theta_2,\theta_3)_{2-+}$ gives the heavy breather
formfactor $F_2$. The corresponding $\sin$ formfactor has no
annihilation poles.
In the soliton-antisoliton-odd breather sector we obtain
\bea
F^{cos}(\theta_1,\theta_2,\theta_3)_{-+1} &=&
-2\pi \frac{(2d)^2 \sqrt{Z}}{2^\frac{3}{2}3^\frac{1}{4}}
\frac{e^{-\frac{1}{2}(\theta_1+\theta_2)-\theta_3}}{\sinh\frac{3\theta_{21}}{2}
\sinh 3\theta_{31} \sinh 3\theta_{23}}\ \zeta(\theta_{12})
\zeta(-\ri\frac{2\pi}{3}) \zeta(\theta_{13}+\ri\frac{\pi}{3})\nn
&\times&
\zeta(\theta_{13}-\ri\frac{\pi}{3})
\zeta(\theta_{23}+\ri\frac{\pi}{3})
\zeta(\theta_{23}-\ri\frac{\pi}{3})
[e^{\theta_1}+e^{\theta_2}+e^{\theta_3}]\nn
&\times& [e^{-\theta_1}+e^{-\theta_2}+e^{-\theta_3}]
[e^{\theta_1+\theta_2}+e^{\theta_3}(e^{\theta_1}+e^{\theta_2})+e^{2\theta_3}]
\nn
&=&\ri\coth\frac{3\theta_{21}}{2}\
F^{sin}(\theta_1,\theta_2,\theta_3)_{-+1}\ .
\label{pm1}
\eea
From the residues at the poles of (\ref{pm1}) we can derive the
breather-breather formfactors:
\bea
F^{cos}(\theta_1,\theta_2)_{11} &=&
-2\pi \frac{4(2d)^2 \sqrt{Z}}{3^\frac{3}{2}}\frac{(\cosh\frac{\theta_{12}}{2})
^2
[\cosh\theta_{12}+\frac{1}{2}]}{(\sinh 3\theta_{12})^2}\nn
&\times&
\zeta^2(\theta_{21})\zeta^2(-\ri \frac{2\pi}{3})
\zeta(\theta_{21}+\ri\frac{2\pi}{3})
\zeta(\theta_{21}-\ri\frac{2\pi}{3})
S_0(\theta_{21}+\ri\frac{\pi}{3})S_0(\theta_{21}-\ri\frac{\pi}{3})\ .
\eea
This is identical to the soliton-antisoliton formfactor
$F^{cos}(\theta_1,\theta_2)_{-+}$ as can be proved by direct
calculation. Some useful identities are 
$S_0(\theta+\ri\frac{\pi}{3})S_0(\theta-\ri\frac{\pi}{3})=S_0(\theta)$,
$2\zeta(\theta)\zeta(\theta-i\pi)=\frac{\sinh 3\theta}
{\sinh\theta+i\sin\frac{\pi}{3}}$ and
\be
\exp\left(\int_0^\infty\frac{dx}{x}\frac{\cosh\frac{x}{6}-\cosh\frac{x}{2}}
{\sinh x\cosh\frac{x}{2}}\right)=\frac{2\sqrt{2}}{c}\ .
\ee
The other breather-breather formfactors are given by
\bea
F^{cos}(\theta_1,\theta_2)_{22} &=&
-2\pi \frac{4(2d)^2 \sqrt{Z}}{\sqrt{3}}\frac{(\cosh\frac{\theta_{12}}{2})^2
[\cosh\theta_{12}+\frac{3}{2}]}{(\cosh 3\theta_{12})^2}\nn
&\times& \zeta^2(\theta_{12})\zeta^2(-\ri \frac{\pi}{3})
\zeta(\theta_{12}+\ri\frac{\pi}{3})\zeta(\theta_{12}-\ri\frac{\pi}{3})\
,\nn
F^{sin}(\theta_1,\theta_2)_{21} &=& -2\pi \frac{4(2d)^2
\sqrt{Z}}{3^\frac{3}{2}}\frac{(\cosh\theta_{12}+\frac{\sqrt{3}}{2}) 
[1+\frac{\sqrt{3}}{2}\cosh \theta_{12}]}{(\cosh 3\theta_{12})^2}\nn
&\times& \zeta(-\ri \frac{\pi}{3})\zeta(-\ri \frac{2\pi}{3})
\zeta(\theta_{12}+\ri\frac{\pi}{2})\zeta(\theta_{12}-\ri\frac{\pi}{2})
\zeta(\theta_{12}+\ri\frac{\pi}{6})\zeta(\theta_{12}-\ri\frac{\pi}{6})\
.
\eea
The special values of $\zeta$ at the breather poles are given by
$\zeta(-\ri\frac{\pi}{3})\approx -1.10184 \ri$ and
$\zeta(-\ri\frac{2\pi}{3})\approx -2.72272 \ri$. We note that
$\zeta(-\ri\frac{\pi}{3})\zeta(-\ri\frac{2\pi}{3})=-3$.

It is apparent that $n$-particle formfactors depend only on $n-1$
independent rapidity variables. This fact can be used to essentially
simplify expressions like \r{dcos} for correlation functions. For
example, 2-particle contributions for $\omega \geq 0$ are given by
\bea
D^{cos}(\omega,q)\bigg|_{\rm 2-part.}&=&- \int_0^\infty 
\frac{\rd\theta}{\pi}\left\lbrace
\frac{2|F^{cos}(\theta)_{+-}|^2 + |F^{cos}(\theta)_{11}|^2}
{s^2-4M^2\cosh^2\frac{\theta}{2}+i\eps}+
\frac{|F^{cos}(\theta)_{22}|^2}{s^2-12M^2\cosh^2\frac{\theta}{2}+i\eps}
\right\rbrace,\nn
D^{sin}(\omega,q)\bigg|_{\rm 2-part.}&=&- \int_0^\infty 
\frac{\rd\theta}{\pi}\left\lbrace
\frac{2|F^{sin}(\theta)_{+-}|^2}
{s^2-4M^2\cosh^2\frac{\theta}{2}+i\eps}
+\frac{2|F^{sin}(\theta)_{21}|^2}
{s^2-4M^2(1+\frac{\sqrt{3}}{2}\cosh\theta)+i\eps}\right\rbrace
\eea
where we also have made use of various symmetry properties of the
formfactors in order to perform the sum over $\epsilon_j$.
Similarly the contribution of formfactors involving one soliton, one
antisoliton and one (light) breather of type 1 can be brought to the
form 
\be
-\int_{-\infty}^{\infty}\frac{\rd\theta}{2\pi}
\int_{-\infty}^{\infty}\frac{\rd\theta_{12}}{2\pi}\ 
\frac{2}{s^2-M^2(1+4\cosh\theta\cosh\frac{\theta_{12}}{2}
+[2\cosh\frac{\theta_{12}}{2}]^2)}|F^{cos}(\theta,\theta_{12})_{-+1}|^2\ ,
\ee
where
\bea
F^{cos}(\theta,\theta_{12})_{-+1}&=&-2\pi \frac{(2d)^2 \sqrt{Z}}
{2^\frac{3}{2}3^\frac{1}{4}}\frac{(2\cosh\t12h+e^\theta)
(2\cosh\t12h+e^{-\theta})(2\cosh\t12h+2\cosh\theta)}{\sinh 3\t12h 
\sinh 3(\theta+\t12h) \sinh 3(\theta-\t12h)} \zeta(\theta_{12})
\zeta(-\ri\frac{2\pi}{3})\nn
&\times&
\zeta(\t12h-\theta-\ri\frac{\pi}{3})\zeta(\t12h-\theta+\ri\frac{\pi}{3})
\zeta(\t12h+\theta-\ri\frac{\pi}{3})\zeta(\t12h+\theta+\ri\frac{\pi}{3}).
\eea

\end{document}